\documentclass[a4paper,12pt]{article}  
\usepackage[T1]{fontenc} 
\usepackage{lmodern}
\usepackage[utf8]{inputenc}
\usepackage[english]{babel}
\usepackage{amsmath}
\usepackage{amssymb}
\usepackage{multirow}

\usepackage{amsthm}
\usepackage{enumerate}
\usepackage{amscd}
\usepackage{mathrsfs}
\usepackage[all]{xy}
\usepackage{graphicx}
\usepackage{color}
\usepackage{tikz}
\usepackage{tocvsec2}

 \usepackage{amsmath, amssymb, amsfonts, graphicx, pdflscape, mathtools, amsthm, upgreek,bm,pgfplots,csvsimple,multirow, multicol, booktabs}

\usepackage{needspace}
\usepackage{tocvsec2}
\allowdisplaybreaks[1]

\usepackage{setspace} 
\usepackage[vmargin = 1in, hmargin =1in]{geometry}
\let\oldfootnote\footnote
\renewcommand{\footnote}[1]{%
  \oldfootnote{\linespread{1.0}\selectfont #1}%
}

\usepackage[authoryear, round]{natbib}
\usepackage{pstool}
\usepackage{accents}
\usepackage{caption}
\usepackage{subcaption}

\usepackage{authblk}

\pgfplotsset{compat=1.10}
\usepgfplotslibrary{fillbetween}
\usetikzlibrary{patterns}

\usepackage{mathtools}

\definecolor{Blue}{RGB}{86,180,233}
\definecolor{Orange}{RGB}{230,159,0}
\definecolor{Green}{RGB}{0,158,115}
\definecolor{DarkBlue}{RGB}{42, 93, 176} 
\usepackage[
	pagebackref,
	colorlinks=true,
	citecolor= DarkBlue,
	linkcolor=DarkBlue,
	urlcolor = DarkBlue
]{hyperref}
\usepackage{cleveref}

\DeclareMathOperator*{\argmax}{arg\,max}

\def\g{\gamma}

\def\e{\varepsilon}

\def\th{\theta}

\def\l{\lambda}

\def\t{\tau}

\def\w{\omega}

\def\D{\Delta}

\def\ol{\overline}

\DeclareMathOperator{\E}{\mathbb{E}}

\newcommand{\paren}[1]{( #1 )} 

\newcommand{\Paren}[1]{\left( #1 \right)}

\newcommand{\Brac}[1]{\left[ #1 \right]}

\newcommand{\de}{\mathop{}\!\mathrm{d}}

\newenvironment{dense}{
 \medmuskip=2mu
 \thinmuskip=0mu
 \thickmuskip=4mu}

\theoremstyle{plain}

\newtheorem{lemma}{Lemma}
\newtheorem{proposition}{Proposition}
\newtheorem{corollary}{Corollary}

\newtheorem{observation}{Observation}

\newtheorem*{example*}{Example}

\theoremstyle{definition}

\newtheorem*{definition*}{Definition}

\usetikzlibrary{positioning}
\usetikzlibrary{snakes}
\usetikzlibrary{calc} 
\usetikzlibrary{arrows}
\usetikzlibrary{decorations.markings}
\usetikzlibrary{shapes.misc}
\usetikzlibrary{matrix,shapes,arrows,fit,tikzmark}
\usetikzlibrary{cd}
\tikzset{
  solid node/.style={circle,draw,inner sep=1.2,fill=black},
  hollow node/.style={circle,draw,inner sep=1.2},
}

\usepgfplotslibrary{groupplots}
\pgfplotsset{compat=newest}

\title{Dynamic Evidence Disclosure: \\
Delay the Good to Accelerate the Bad\thanks{We are grateful to Yingni Guo for an excellent discussion. We thank Ian Ball, Simon Board, Henrique Castro-Pires, Kailin Chen, Inga Deimen, Ilan Guttman,  Daniel Hauser, Johannes H\"{o}rner, Deniz Kattwinkel, Tuomas Laiho, George Mailath, Pauli Murto, Ludovic Renou, Ali Shourideh, Vasiliki Skreta, Andy Skrzypacz, Peter Norman S\o rensen, Egor Starkov,  Juuso Toikka, Juuso V\"alim\"aki, Leeat Yariv, and Weijie Zhong, as well as audiences at ASU, U of Arizona, BC,
Berlin IO Day, CETC, Columbia,  Copenhagen, EC '24, LSE, Miami, MSET, Nordic Theory Conference, Northwestern, Oxford, UPenn, Pittsburgh-CMU, PSE, QMUL-City Theory WS, RHUL, Rice,  SAET, SITE, Surrey, Theory in Rio, TSE, UB, Warwick Theory Conference, WiET, WUSTL, and Yale for their insightful comments.}  
}

\author{\href{https://sites.google.com/view/knoepfle}{Jan Knoepfle}\thanks{\linespread{1.1}\selectfont School of Economics and Finance, Queen Mary University of London,  j.knoepfle@qmul.ac.uk.} \
  and \href{https://sites.google.com/view/juliasalmi}{Julia Salmi}\thanks{\linespread{1.1}\selectfont  Hanken School of Economics and Helsinki GSE, julia.salmi@hanken.fi.} 

 }

\begin{document}

\begin{titlepage}

\maketitle \thispagestyle{empty} 

\begin{abstract}
\linespread{1.1}\selectfont
We analyze the dynamic tradeoff between generating and disclosing evidence.
Agents are tempted to delay investing in a new technology in order to learn from information generated by the experiences of others. This informational free-riding is collectively harmful as it slows down learning and innovation adoption. 
A welfare-maximizing designer can delay the disclosure of previously generated information in order to speed up adoption. 
The optimal policy transparently discloses bad news and delays good news. 
This finding resonates with regulation demanding that fatal breakdowns be reported promptly. 
The designer's intervention makes all agents better off. 
\end{abstract}
\noindent
{\linespread{1.1}\selectfont Keywords: innovation, social experimentation, dynamic disclosure, evidence}

\end{titlepage}


\section{Introduction}
Society is constantly presented with opportunities to adopt new technologies, from groundbreaking medical treatments to innovative devices or software. 
Early stages of new technologies carry significant uncertainty about quality or viability. 
This uncertainty is reduced only as agents adopt and experience the new technology. 
However, the information generated by early adopters only benefits those who choose later, leading agents to adopt new technologies at a slower pace than socially optimal, slowing down learning and reducing potential benefits to society.\footnote{Take the case of COVID-19 vaccines. Surveys indicated that only 30\% planned to get a vaccine soon after becoming available \citep[see][]{Schaffer2020Vaccination}.}

This paper studies the dynamic tradeoff between evidence \textit{generation} and \textit{disclosure}.   
To explore the optimal balance, we introduce a designer who controls what evidence is disclosed at what time and analyze welfare-maximizing policies. 
Conditional on evidence having been generated, welfare is maximized by disclosing it immediately. However, immediate disclosure creates incentives for informational free-riding as agents prefer to wait, which is inefficient when generation is endogenous.
By committing to delay information disclosure, the designer can mitigate inefficient free-riding. The designer faces a tradeoff as fast disclosure enables better decisions, but it also increases the agents' value of waiting, slowing down information generation. Importantly, this tradeoff is dynamic: committing to delay \textit{future} disclosures incentivizes investment \textit{today}.

In an environment with bad and good news, we show that different types of news affect agent-behavior on different margins. Bad news is decision relevant on the \textit{action margin} -- whether or not to invest. Good news is relevant on the \textit{timing margin} -- when to invest.
Any optimal policy delays good news to encourage faster evidence generation and discloses bad news immediately to prevent inefficient investments. Hence,  optimal disclosures delay information on the timing margin in order to improve the action margin.

We propose 
a game in continuous time between a designer and a continuum of agents. Each agent chooses the time at which he makes an irreversible investment, if ever. Agents differ in discount rates and their capabilities to generate evidence. Each agent's net payoff at the time of investment is positive or negative, depending on an unknown state of the world, which can be good or bad. A fully revealing piece of evidence about the state is generated at rates proportional to the mass of new investments.

The designer commits to a public disclosure policy to maximize the expected discounted welfare of the agents. 
A disclosure policy specifies when and what type of evidence to disclose. Evidence is hard in that prior generation is necessary for disclosure, the designer cannot fake evidence. 
For a non-trivial problem, some agent must be willing to invest first, so we assume that the prior about the state is high enough that immediate investment is optimal myopically, i.e. if no evidence was ever disclosed.

Considering the agents' investment choices under a given policy, note that optimal behavior after the disclosure of fully revealing evidence is trivial: all agents who have not invested previously invest immediately after good news and never invest after bad news. 
Non-trivial investment dynamics occur while no evidence has been disclosed yet. 
 We say an agent \textit{experiments} if he invests before observing evidence.
 In any equilibrium, all less patient agents must have invested before more patient agents start to experiment.

To understand how bad and good news, and the different margins they affect, shape the optimal policy, we study each in isolation first.
In \Cref{sec:pureBad}, we analyze the case in which only bad news can be generated. 
Under pure bad news, the only inefficiency is that experimentation is too slow compared to the first best. Since the absence of disclosure is good news in this case, agents become gradually more optimistic and everyone eventually experiments if no evidence is disclosed. However, each agent prefers others to bear the experimentation cost of generating evidence, so they are willing to experiment only if bad news is not disclosed too quickly, this free-riding delays experimentation. 

We argue that this inefficiency cannot be corrected by delaying disclosures. Consider an agent's expected payoff from experimenting at time $t$. Holding fixed the probability that bad evidence is disclosed before $t$, this payoff is independent of how long before $t$ the agent learns that he will not invest. Hence, bad news is not decision-relevant on the \emph{timing margin}. By contrast, increasing the probability that bad evidence is disclosed by time $t$ raises the payoff from experimenting at $t$ because bad news is decision-relevant on the \emph{action margin}; agents experiment only if it is myopically optimal, and observing bad news prevents mistaken investment.
Accordingly, \Cref{prop:OptimalpureBN} establishes that it is impossible at any time $t$ to raise the probability of disclosing evidence beyond that under transparency. Thus, although delaying disclosure can speed up evidence \emph{generation},\footnote{At the extreme, if no evidence is ever disclosed, all agents invest immediately.} it is not possible to \emph{disclose} the additional information at any time without inducing some agents to delay experimentation. Transparency is optimal under pure bad news.

\Cref{sec:pureGood} considers the opposite extreme where only good news can be generated. 
Good news is decision-relevant only on the {timing margin} while agents experiment as evidence does not alter the myopically optimal action. The \emph{lack of good news} can be action-relevant if it is sufficiently bad news so that agents are unwilling to experiment.
With pure good news, the inefficiency arises from the amount of experimentation rather than its speed. Under both transparency and the first best, all agents who experiment do so at time $t=0$. Since no agent experiments with negative expected return, experimentation under transparency stops before the lack of good evidence becomes action-relevant.

 The optimal (incentive-compatible) policy increases welfare through disclosure delays which induce an action-relevant amount of experimentation. 
\Cref{prop:OptimalPureGN} shows that whenever pure good news is relevant for welfare, all generated evidence must be delayed until a single disclosure time.
Because the opportunity cost of experimentation under the optimal policy -- delayed investment if others generate evidence -- is lower than under the first best -- immediate investment in case of evidence -- the amount of experimentation under the optimal policy exceeds the first-best level if the potential for good news generation is high (see \Cref{prop:amountcompNEW3}).

Finally, \Cref{sec:optimal_policy} considers the general case where both good and bad evidence can be generated. 
\Cref{prop:main2NEW} shows that any optimal policy must combine the key features of the pure-news cases: all bad evidence generated by more impatient agents must be disclosed before any more patient agent experiments, and all good evidence must be delayed at least until the most patient agents who ever experiment begin to do so. 
An important difference is that delaying good news always improves welfare unlike under pure good news. 
\Cref{prop:main1NEW} then identifies a particularly simple optimal policy: disclose bad news transparently as soon as generated and delay all good news until a single time. 

With both types of evidence, delaying good news is always necessary because it accelerates bad news. 
The optimality of delaying good news stems from the distinct margins on which good and bad evidence act. The rate at which a delay of good news accelerates bad news depends on the discount rate of the marginal agent who generates it. The remaining agents value avoiding the wrong action more than investing early, so they benefit when action-relevant bad news is accelerated by postponing timing-relevant good news. Conversely, earlier (less patient) agents are not induced to wait longer.

The optimal disclosure policy yields a Pareto improvement over transparency, and now both the amount and rate of experimentation are higher in the second best than under transparency. 
To conclude the treatment of the general case, we analyze conditions under which the designer's intervention generates the largest welfare gains and describe the resulting adoption dynamics under optimal disclosure.

Endogenous learning through sufficient experience of early adopters is a crucial factor in the roll-out of any new technology\footnote{PwC's ``Global Artificial Intelligence'' study estimates that AI could contribute up to \$15.7 trillion to the global economy by 2030. 
 The UK government's 2023 white paper on artificial intelligence (AI) states ``[I]ndustry repeatedly emphasised that consumer trust is key to the success of innovation economies.'' and 
 ``By building trust, we can accelerate the adoption of AI across the UK to maximise the economic and social benefits.'' See 
https://pwc.co.uk/industries/financial-services/understanding-regulatory-developments/ai-in-financial-services-navigating-the-risk-opportunity-equation.html and https://gov.uk/government/publications/ai-regulation-a-pro-innovation-approach/white-paper.}
or practice, for example elective medical treatments or legislative changes in different states and countries.\footnote{Consider the 2021 decriminalization of hard drugs in Oregon (Measure 110). After an increase in drug trade and related crimes, other states became less enthusiastic about adopting similar policies.}
When planning for a novel vaccination program, public hesitancy is a major concern.
The optimal policy shares features of so-called ``confirmatory clinical trials with intention to treat.'' These trials verify the efficacy and safety of a new treatment during a pre-specified horizon at the end of which the collected evidence is published to validate the benefits. 
Adverse effects, however, have to be reported immediately. 
Our findings support such reporting mandates: even though the prospect of prompt reporting may encourage individuals to postpone, revealing bad news is without loss in terms of social learning.\footnote{There are natural reasons to immediately report adverse effects that are beyond the scope of our model. The duty of care owed by medical professionals would likely impede censoring indications of a health threat to accelerate testing. Our results lend additional support to such transparency rules.}

\subsection{Related literature}

This paper is related to the literature on dynamic disclosure, including 
dynamic persuasion of myopic agents \citep{ely2017,renault2017optimal,arieli2024}, and forward-looking agents \citep{ball2023, ely2020, orlov2020,zhao2022contracting,escude2023slow},\footnote{For information-design with exogenous generation and a single receiver facing a stopping problem, see \cite{au2015dynamic}, \cite{che2023keeping}, 
\cite{knoepfle2020dynamic}, \cite{hebert2022engagement}, \cite{koh2022attention}, \cite{saeedi2024getting} and \cite{koh2024}.} as well as 
evidence disclosure without commitment \citep{acharya2011,guttman2014,gratton2018,chatterjee2024,zhou2024}. 
The most important difference in our setting is that the information available to the designer is endogenously determined by her disclosure choices, which relates to a recent stream of work we discuss below. 

Another related strand studies social experimentation by forward-looking agents without endogenous disclosures \citep{rob1991,Frick2016,garfagnini2016social,LaihoSalmi2019,SalmiLaihoMurto2020}.\footnote{This work is closely related to observational learning with timing decisions: \cite{chamley1994}, \cite{Murto2011}, and \cite{wagner2018}. More recently, \cite{meyer2025delay} characterizes conditions under which no delay arises when a player can learn from the outcome or the action of his opponent.} The transparent benchmark under pure pure bad news (\Cref{sec:pureBad})  is most closely related to the equilibria in \cite{Frick2016} and \cite{LaihoSalmi2019}.
\cite{Frick2016} show that increasing the potential to \textit{generate} negative evidence beyond a saturation point has no impact on welfare. 
 We show how interventions that speed up the \textit{disclosure} of negative evidence, e.g., delaying good evidence, increase welfare.\footnote{\cite{Frick2016} note that increasing the potential for learning \emph{can} hurt welfare if they considered heterogeneous agents \cite[see Remark 1 on p. 1521]{Frick2016}. This is because adoption opportunities arrive at a bounded rate in \cite{Frick2016}.
 \cite{omaPE} find a negative effect on equilibrium learning under direct network externalities. In a less related environment, \cite{board2024} find that a fixed observation lag may improve experimentation.} 

Our paper is closest to settings that recently attracted attention, where the designer's disclosure choices influence agents' actions and, in turn, what the designer learns.

\cite{halac2017contests} study optimal feedback and contest design, where agents choose each period whether to exert effort.\footnote{See also \cite{goltsman2011interim}, \cite{aoyagi2010information} and \cite{ely2022} for feedback in contests without uncertainty about the environment. \cite{smolin2021} and \cite{ely2023} study feedback to a single agent in dynamic moral hazard. 
\cite{cetemen2020uncertainty} consider blended feedback about quality and aggregate effort in a team-production problem.} They show that in equal-sharing contests -- where, as in our setting, there is no competition among agents -- concealing breakthroughs is optimal to keep agents optimistic about quality.  We 
 identify an additional benefit of hiding breakthroughs when agents' choices are irreversible: reducing the value of waiting. 

The adoption problem in our paper is closer to \cite{kremer2014} and \cite{CheHorner2017}, who study optimal disclosure to persuade a sequence of short-lived agents to experiment.\footnote{See also the more recent works about endogenous learning from short-lived consumers by \cite{fainmesser2023consumer}, where a platform informs sellers' pricing choices;  and by \cite{janssen2025dual}, where a search engine ranks products.} In contrast, agents in our model are forward looking and disclosure is public.\footnote{We discuss private disclosures in Section \ref{sec:discussion}. Contributions on public disclosure to multiple receivers include \cite{laclau2017public} and \cite{inostroza2024adversarial}.} These features lead to substantially different strategic effects and implications.
For instance, with myopic agents -- or when future disclosures can target agents privately -- good evidence should never be delayed, since future disclosure does not slow down today’s evidence generation. 
 The incentive constraints that deter delays also make the welfare-maximizing policy in our setting a Pareto improvement over transparent disclosure; in \cite{CheHorner2017} earlier agents are made strictly worse off to benefit later adopters. 

To the best of our knowledge, the only papers in this stream studying disclosure to long-lived agents who take irreversible actions are the contemporaneous work by \cite{chen2024optimal} and subsequent work by \cite{lyu2025Patient}. \cite{chen2024optimal} analyze information design under heterogeneity in adoption payoffs and a finite deadline and consider pure bad news and pure good news separately. Their designer crucially exploits her freedom to ``fake'' news to reveal partial information. 
Our two papers are complementary. We focus on the \emph{timing} dimension -- where the designer delays the disclosure of hard evidence -- while \cite{chen2024optimal} focus on the optimal design of signals. Finally, \cite{meyertervehn2025} and \cite{koh2025balanced} study optimal information revelation to social learners who move in a fixed order, which makes agents act myopically as in \cite{che2023keeping} and \cite{kremer2014}. 
The fixed order of moves allows them to examine the benefits of alternative informational regimes more broadly.

\section{Model} \label{sec:model}

\noindent
\textbf{Environment.} Time $t \ge 0 $ is continuous. 
There is a continuum of agents indexed by $\theta \in [0,F_n]$ and one designer. 
Each agent chooses the time of an irreversible investment, with investment time $\t=\infty$ if the agent never invests. The designer controls evidence disclosure as described shortly. 
If an agent invests at time $\t$, the game ends for this agent and he collects  payoff $v_{\omega}$, which depends on the unknown state of the world $\omega\in\{G,B\}$. Payoffs satisfy $v_{G}> 0 > v_{B}$ and capture the cost and the benefit of investment. 
The common prior is $x_0 = 
\Pr[\w=G]$. 
The sequence of events in instant $[t, t+\de t)$ can informally be described as follows. First, each agent who remains in the game chooses whether to invest. Then, evidence may be generated. Finally, evidence is disclosed according to the policy. 

\vspace{0.18cm} 

\noindent
\textbf{Evidence generation.} Evidence is generated as agents invest. Let the right-continuous process $q=(q_t)_{t\ge 0}$ denote the cumulative mass of agents who have invested by time $t$.\footnote{We denote left limits by $q_{t-} = \lim_{s\nearrow t} q_s$ and $\dot q_t$ denotes the right derivative of $q$. By convention, $q_{0-}=0$.}    The probability that a piece of fully revealing good evidence has been generated by time $t$ in the good state is $G(q_t)$, and the probability that a fully revealing piece of bad evidence has been generated in the bad state is $B(q_t)$. We assume that $G(0) =B(0) = 0$, and each of them is either constantly zero or strictly increasing and differentiable with densities $g$ and $b$ and everywhere finite hazard rate.
One example used for illustrations and figures below is the exponential technology  with $B(q) = 1- e^{-\l_B q}$ and $G(q) = 1- e^{-\l_G q}$ for hazard rates $\l_G, \l_B \ge 0$. 
For the general technology, the hazard rate $g(q)/(1-G(q))$ determines the potential for good evidence generation by the $q$'th adopter. Going beyond exponential generation allows us, for example, to capture instances where earlier adopters are better at generating information in general, or instances where early adopters are likely to reveal positive new features ($g(q)/(1-G(q))$ large for low $q$) but the adoption by many agents is required to uncover potential faults ($b(q)/(1-B(q))$ low initially and then increasing).

\vspace{0.18cm} 

\noindent
\textbf{Evidence disclosure.}
The designer controls what agents learn by committing to a \emph{public} disclosure policy. We model information as \emph{hard evidence}. The designer can delay or hide evidence, but she cannot manipulate or fabricate it.\footnote{We discuss more flexible information design tools in \Cref{sec:discussion}. In particular, private disclosures would enable the designer to implement the first best.}
Formally, the designer chooses two families of cumulative distribution functions $D^\omega(\cdot  \lvert s)\colon [0,\infty] \to [0,1]$ for $\omega\in\{G,B\}$ and all $s\ge 0$, such that, if $\omega$-evidence is first generated at time $s$, then the cdf $D^\w(\cdot  \lvert s)$ defines the distribution of the delay $\D \ge 0$ until this piece of $\w$-evidence is disclosed.

Agents only care about the time evidence is disclosed, not when it is generated.
We can thus drastically simplify the designer's problem and represent disclosure policies directly in terms of the expected disclosure probabilities, integrating over all possible (i.e. earlier) generation times: the non-decreasing processes $H^\w = (H_t^\w)_{t\ge 0}$ for $\w \in \{G,B\}$. 
Since evidence must be generated before disclosure, any policy must satisfy $H_t^B \le B(q_{t-})$ and $H_t^G \le G(q_{t-})$ for all $t\ge 0$. In Online Appendix \ref{sec:Z-processes} we show the equivalence between specifying $(H^B,H^G)$ subject to this restriction and specifying the above mentioned families of conditional delay distributions $D^B(\cdot  \lvert s)$ and $D^G(\cdot  \lvert s)$ for each generation time $s$.
We define $H^\w$ as left-continuous processes, with the interpretation that $H^\w_t$ denotes the probability that $\w$-evidence is disclosed \textit{strictly} before $t$.\footnote{This way, agents at time $t$ base their actions on $H_t^\w$ and we avoid explicitly referring to the left-limit in all expressions.}

\vspace{0.18cm} 

\noindent
\textbf{Updating.}
Once good (resp. bad) evidence is disclosed, the agents' belief jumps to 1 (resp. 0). 
While no evidence has been disclosed, the agents update the belief about the state taking into account the disclosure policy.  
The no-disclosure belief at time $t$ is 
\begin{align}\label{belief-Q}
	x_t= \frac{x_{0}(1-H_t^G)}{x_{0}\Paren{1-H_t^G}+(1-x_0)\Paren{1-H_t^B}}, 
\end{align}
where $x_{0}(1-H_t^G)$ is the probability that the state is $G$ and no disclosure and $(1-x_0)(1- H_t^B)$ is the probability that the state is $B$ and no disclosure.
The no-disclosure belief increase or decrease at $t$, depending on whether $\frac{1-H^G_t}{1- H^B_t} $ increases or decreases at $t$. 
Let $x(\th):=\frac{x_{0}(1-G(\th))}{x_{0}(1-G(\th))+(1-x_0)(1-B(\th))}$ be the belief based on the knowledge that no news was \textit{generated} after agents in $[0,\th]$ experimented.

\vspace{0.18cm} 

\noindent
\textbf{Payoffs.}
The agent's index $\th \in [0, F_n]$ determines his discount rate: $r(\th) \in \{r_1, \cdots, r_n\}$. Without loss, we order agents and discount rates such that $r_i > r_{i+1}$ for all $i$ and $r(\th) \ge r(\th')$ whenever $\th \le \th'$, i.e. agents with higher $\th$ are more patient. 
Let $f_i$ denote the mass of agents with discount rate $r_i$ and define $F_i = \sum_{j=1}^i f_j$.\footnote{Formally, given the increasing mapping $r
$, $F_i = \sup\{\th: r(\th)\le r_i\}$ and $f_i = F_i - F_{i-1}$, with the convention $F_0 = 0$. While either the mapping $r$ or the distribution $F$ would suffice to specify the setup, having both objects drastically simplifies the statements. } 
 Let $v(x) = x v_G +(1-x)v_B$ denote the expected value from investing at belief $x$. 
In the absence of additional information, agents would invest immediately if and only if the belief is above the \textit{myopic threshold} $x^\text{myop}:=\frac{-v_B}{v_G-v_B}$. 
 Assume the prior is above the threshold: $x_0>x^\text{myop}$.\footnote{\linespread{1.0}\selectfont Without this assumption, no investment can be induced under any disclosure policy.}

Consider the problem of agent $\th$ under some policy $(H^B,H^G)$. 
The optimal choice upon disclosure is trivial: invest immediately if good and never invest if bad news is disclosed. 
Absent disclosure, taking as given the processes $H^\w$, agent $\th$ solves 
  \begin{dense}
\begin{align}\label{eq:agent_prob}
    \sup_{\t \in\,[0,\infty]}  \; \bigg\{ \int_0^{\t} e^{-r(\th) t} x_0 v_G \de H^G_t   + e^{-r(\th) \t }
  \Paren{x_0 v_G (1-H^G_\t ) + (1-x_0)v_B (1-H^B_\t)} \bigg\}. 
\end{align}
 \end{dense}
The first term captures the event that good evidence is disclosed at some time $t\le \t$, i.e. before the agent experiments absent disclosure.  
 The second term corresponds to the case of no disclosure by time $\t$.  It is easily verified that the term in parentheses equals $\paren{x_0 (1-H^G_{\t} ) + (1-x_0) (1-H^B_{\t})}\,v(x_{\t})$, the probability that no evidence was disclosed times the conditional expected investment return.  
For values of $\tau$ that are not strictly dominated by $\tau' =\infty$, the payoff in \eqref{eq:agent_prob} satisfies the single-crossing property in $(\tau, r)$,\footnote{Since the integrand in \eqref{eq:agent_prob} is positive, stopping at $\tau$ with  $x_0 v_G (1-H^G_{\t} ) + (1-x_0)v_B (1-H^B_{\t})<0$ is strictly dominated by stopping at $\t'=\infty$. Single-crossing follows immediately for the equivalent objective which considers only the positive part $\Paren{x_0 v_G (1-H^G_{\t})+ (1-x_0)v_B (1-H^B_{\t})}^+$.}  which implies that more impatient agents invest weakly earlier than more patient agents under any disclosure policy. Thus, assume (wlog) throughout that agents invest in order of $\th$ so that $\th = q_t$ is the index of the last agent who has experimented at $t$.

\subsection{The designer's problem} \label{sec:prin_problem}
We solve for the disclosure policy that maximizes the time-0 expected welfare. 
A disclosure  policy specifies two non-decreasing processes $ (H^G, H^B)$. %
Given a policy, we call a process $q$ that is consistent with agents playing best responses \textit{incentive compatible}: 
\begin{definition*}
    Given a policy $(H^G, H^B)$, the experimentation path $q$ is \emph{incentive compatible} if for all $\th$, the time $\t(\th) := \inf\{ t \colon q_t \ge \th \}$ satisfies
    \begin{dense}
    \begin{align*}
        \t(\th) \in \argmax_{\t \in [0,\infty]} &&{\Big\{}   \int_0^{\tau} e^{-r(\th) s} x_0 v_G \de H^G_s   + e^{-r(\th) \t } \Paren{x_0 v_G (1-H^G_{\t})  + (1-x_0)v_B (1-H^B_{\t})} {\Big \}}. 
    \end{align*}
        \end{dense}
\end{definition*}
Incentive compatibility requires the $\argmax$ to be non-empty for all $\th$. Thus, for some policies no incentive compatible $q$ exists. 
Conversely, optimal choices may lead to a $q$ under which the policy is infeasible, e.g. if $H^G_t > G(q_t)$ for some $t$. Thus, we also require:

\begin{definition*}
    A policy $(H^G, H^B)$ is \emph{feasible} given $q$ if they jointly satisfy  
    \begin{align*}
     H^G_t \le G(q_{t-}) \quad \quad \text{and} \quad \quad H^B_t \le B(q_{t-}), \quad  \quad  \quad \text{for all times } t.  
    \end{align*}
\end{definition*}
Since $G(q_{t})$ is right-continuous and non-decreasing, $H^G_{t} \le G(q_{t-})$ for all $t$ implies $H^G_{t+} \le G(q_{t})$ for all $t$,\footnote{Suppose $H^G_{t+} > G(q_{t})$ for some $t$. Then, for sufficiently small $\e>0$ we have $H^G_{s} - G(q_{s}) > 0$ for all $s\in (t,t+\e]$.
Since $G(q_{t})$ is increasing and $q_{s-}\le q_{s}$, we get $H^G_s - G(q_{s-}) > 0$ for $s\in (t,t+\e]$. This would violate the feasibility constraint at $t+\e$.} similar for $H^B_{t+} \le B(q_t)$.
A policy $(H^B,H^G)$ is \textit{implementable} if there is an incentive compatible $q$ such that $(H^B,H^G)$ is feasible given $q$.

The designer's problem is 
\begin{dense}  
\begin{multline}\label{eq:MaxProblem_general}
\begin{aligned}
    &\underset{H^B, H^G, q}{\sup}  && \Big \{ \int_{0}^{F_n} \Big [ \int_0^{\tau(\th)} e^{-r(\th) s} x_0 v_G \de H^G_s 
    \\
    & &&\hphantom{ \Big \{ \int_{0}^{F_n} \Big [ \int_0 } + e^{-r(\th) \tau(\th) }\Paren{x_0 (1-H^G_{\tau(\th)} ) v_G + (1-x_0)(1-H^B_{\tau(\th)} )v_B}  \Big ]\de \th \ \Big \},
    \\
    & \text{such that } &&  (H^B,H^G) \text{ is feasible given } q,
     \\
    & \text{and }
 && q  \text{ is incentive compatible given } (H^B,H^G). 
\raisetag{5em}
   \end{aligned}
\end{multline}
\end{dense}

We refer to $(H^B,H^G)$ that solves \eqref{eq:MaxProblem_general} as an {optimal policy}. 
 Before solving the problem for the general model with joint good- and bad-news learning in Section \ref{sec:optimal_policy}, we study the special cases of pure bad news and pure good news separately.

\section{Pure bad-news learning}\label{sec:pureBad}
Consider pure bad news, i.e., $ H^G \equiv G \equiv 0$ throughout this section. We will show that transparent disclosure of bad news is optimal.

\subsection{Transparent benchmark under pure bad news}
Consider the equilibrium under transparent disclosure, that is $H^B_t = B(q_{t-})$ for all $t$.\footnote{A formal derivation of the equilibrium under transparent disclosure with joint bad and good news, which covers the following as a special case, can be found in \Cref{Asec-TP}.}
Let $q_t^\text{TP}$ denote the mass of agents who have experimented by time $t$ under transparency. 
With transparent bad news,  agents experiment gradually, so the process $q_t^\text{TP}$ is continuous, and the rate of investment $\dot{q}_t^\text{TP}$ must make the marginal agent $\th = q_t^\text{TP}$ indifferent between experimenting and waiting for an additional instant:\footnote{Note that $x_t$ is itself a function of $q^\text{TP}_t$, so the indifference condition \eqref{eq:PureBN_qTB} is a first-order ordinary differential equation (ODE). 
 The solution to this ODE is provided in \Cref{Asec-TP}.}
\begin{align}\label{eq:PureBN_qTB} 
  r(q^\text{TP}_t) v(x_t)=  (1-x_t) \frac{b(q^\text{TP}_t)}{1-B(q^\text{TP}_t)} \dot q^\text{TP}_t (-v_B)  . 
\end{align} 
The left-hand side gives the cost of waiting per unit of time. The right-hand side gives the informational gain per unit of time: the gain consists in avoiding the negative payoff $v_B$ in case bad evidence arrives, which occurs at rate  $(1-x_t) \frac{b(q^\text{TP}_t)}{1-B(q^\text{TP}_t)} \dot q^\text{TP}_t$.

\begin{figure}[ht]
  \centering

\begin{tikzpicture}
	 \begin{axis}[
                        axis lines = center,
                        scale = 0.75,
                        xtick = {0.001,.192,.284},
                        xticklabels = {$0$,${T_1}$,${T_2}$},
                        ytick = {0.001,.5,1},
                        yticklabels = {$0$,$F_1$,$F_2$},
                        xmin = 0,
                        xmax = .5,
                        ymin = 0,
                        ymax = 1.2, 
                        xlabel = {$t$},
                        ylabel = {$q$
                        },
                        x label style={at={(current axis.right of origin)},anchor=west},    
                        y label style={at={(current axis.above origin)},anchor=south},    
                    ]
                    
                 \addplot [dashed, gray, domain=0:.45, samples=2]{1} ;   
                    \addplot [dashed, gray, domain=0:.192, samples=2]{.5} ;

                              \draw[-,  blue, dashed, thick] (axis cs:0.048,0.056) to  (axis cs:0.495,0.056);
            \fill[blue] (axis cs:0.1625,.056) circle (0.5pt);
                          \fill[blue] (axis cs:0.175,.056) circle (0.5pt);
                 \fill[blue] (axis cs:0.1875,.056) circle (0.5pt);


                              \draw[-,  blue, dashed, thick] (axis cs:0.144,0.26) to  (axis cs:0.49,0.26);

                              \draw[-,  blue, dashed, thick] (axis cs:0.238,0.662) to  (axis cs:0.5,0.662);

                              \draw[-,  blue, dashed, thick] (axis cs:0.275,0.9) to  (axis cs:0.5,0.9);
                              
                              \addplot [very thick, domain=0:.192, samples=50]{1/3*ln(1/(2-exp(3*x)))};

                       \addplot [very thick, domain=.192:.284,samples=50]{ 1/3*ln(1/(2-exp(0.576 + x- .192)))}
                       node[left,pos=.275] {$q_t^\text{TP}$}
                       ; 
                    \addplot [very thick, domain=.2844:.5,samples=2]{1}; 
                    \end{axis}
		\end{tikzpicture} 
  
\caption{\linespread{1.0}\selectfont \small Experimentation amount with $B(q) = 1-e^{-\l_B q}$. At all times prior to $T_2$, negative evidence arrives at a positive rate. If negative evidence arrives, there is no further investment (dashed blue lines). }
\label{fig:PureBN_transp}
\end{figure}

Figure \ref{fig:PureBN_transp} depicts the experimentation  dynamics under transparent disclosure for an example with $n=2$ and constant hazard rate $b(q)/(1-B(q)) = \l_B$. Between times $0$ and $T_1$, the impatient types $r_1$ experiment. As no news is good news when $G=0$, agents become gradually more optimistic. To keep agents with the same discount rate indifferent between experimenting and waiting, the rate of evidence arrival -- and in this example with constant hazard rate, the rate of experimentation -- increases. 
At time $T_1$, all agents with type $r_1$ have experimented and the more patient agents with type $r_2$ start experimenting. Due to the lower cost of waiting ($r_2<r_1$), the experimentation rate of the patient agents is initially lower and increases as the belief increases further. 

With pure bad news, all agents  experiment eventually ($\lim_{t\to\infty} q^\text{TP}_t = F_n$), but experimentation is inefficiently slow. Since the amount of evidence $B(q)$ is determined by solely by how many agents have experimented, all delays are inefficient and caused by strategic free-riding incentives. If the designer could choose agents' strategies, the first best would have them experiment at infinite speed (see \Cref{sec:FBdetailed} for details).

To see the potential benefit of censoring bad news disclosures, suppose the designer was allowed to disclose information privately to each agent. Then, she could implement the first best by disclosing to each agent $\th$ exactly the information that was generated by his predecessors $[0,\th)$ but nothing from other agents. This would completely eliminate the incentive to delay experimentation and inform all agents efficiently. 
This observation raises the question of whether the designer, restricted to public disclosures, can increase welfare by delaying bad news from earlier experimenters to speed up experimentation and benefit later adopters. We now show that the answer is no. 

\subsection{Optimal disclosure of pure bad news}
Transparent disclosure is optimal under pure bad news; the designer maximizes welfare by committing to disclose all bad evidence immediately. This follows from a stronger result, that delaying bad news at any earlier time does not increase the implementable cumulative disclosure probability at any later time. The following result makes reference to the experimentation amount $q^\text{TP}$ under transparency defined by the solution to \eqref{eq:PureBN_qTB}. 

\begin{proposition}\label{prop:OptimalpureBN}
Consider pure bad news and let $\ol H^B_t = B(q^\text{TP}_t)$ be the disclosure process under transparency. 
Any implementable disclosure policy satisfies $H^B_t \le \ol H^B_t$ for all $t$.
\end{proposition}

Optimality of transparency for welfare follows immediately. Increasing $H^B_t$ pointwise increases the agent's payoff \eqref{eq:agent_prob} for all $(r,\t)$. Thus, increasing disclosure from any $H^B$ to $\ol H^B$ and having agents re-optimize can only increase the objective in \eqref{eq:MaxProblem_general}: 
\begin{corollary}\label{cor:PureBNTrans}
Transparent bad news $H^B_t = B(q_t)$ with $q_t = q^\text{TP}_t$ solves problem \eqref{eq:MaxProblem_general}.     
\end{corollary}

\Cref{prop:OptimalpureBN} implies that transparency of bad news is also optimal under different informational objectives. Assume the designer attaches an over-proportional value to bad news being disclosed at specific moments in time. No matter which times are more important, transparency at all times is optimal. 
On the other hand, \Cref{prop:OptimalpureBN} does not imply that any implementable policy induces experimentation paths with $q_t \le q_t^\text{TP}$. By restricting the disclosure of bad evidence, the designer can implement faster information generation than under transparency. For instance, if the designer wanted to maximize adoption at time $t=0$, he could commit never to disclose any information and have all agents adopt immediately. 
\Cref{prop:OptimalpureBN} shows that none of the additionally generated information can be disclosed earlier than under transparency. The proof in \Cref{proof:OptimalBN} shows that doing so would necessarily induce some agents to delay experimentation.

To build intuition, consider an agent who is willing to experiment absent news. If bad news arrives, the agent changes his action from investing eventually to never investing. Thus, bad news is decision relevant on the \emph{action margin}. 
Agent $\th$ is willing to experiment at time $\t$ only if the following holds for all $t\ge \t$:
\begin{align}\label{eq:PureBN_IC}
    e^{-r(\th)\t }\Paren{x_0 v_G + (1-x_0)(1-H^B_\t)v_B } \ge e^{-r(\th) t} \Paren{x_0 v_G + (1-x_0)(1-H^B_{t}) v_B }. 
\end{align} 
The inequality gives the time-0 discounted payoffs from experimenting at time $\t$ and experimenting at time $t$. 
The agent's payoff on the left-hand side depends on the policy $H^B$ only through the value $H^B_\tau$, not on $H^B_s$ for any $s<\tau$.
Thus, we say that bad news is \emph{not} decision relevant on the \emph{timing margin}; conditional on the value of $H^B_\tau$, the exact disclosure time of bad news before $\t$ is irrelevant.
Inequality \eqref{eq:PureBN_IC} puts an upper bound on $H^B_t$ for each $t>\tau$, but this upper bound depends only on the value of $H^B_\tau$ and the period length $t-\t$, not on the value of $H^B$ at any other time. 
  In particular, delaying bad news disclosures from the interval $(\tau,t)$ to $t$ does not make waiting until $t$ less attractive. 
 This will be different for the timing of good-evidence disclosures. 
 
The upper bound $\ol H^B$ in \Cref{prop:OptimalpureBN} can be computed recursively from \eqref{eq:PureBN_IC}: First, set $r(\th)=r_1$ and $\tau=0$. Solve \eqref{eq:PureBN_IC} (binding) for $H^B_t$ for $t\in (0,T_1]$, where $T_1$ is determined by the time at which the resulting solution for $H^B$ satisfies $H^B_{T_1}= B(F_1)$. Next, set $r(\th)=r_2$, $\tau = T_1$, and $H^B_{\t}=B(F_1)$;  and solve  \eqref{eq:PureBN_IC} for $H^B_t$ for $t \in (T_1, T_2]$, with $T_2$ given by $H^B_{T_2}= B(F_2)$. Continuing through all types gives $\ol H^B$.\footnote{It is easily confirmed that solving ODE \eqref{eq:PureBN_qTB} with initial value $q^\text{TP}_0=0$, and setting $\ol H^B_t=B(q^\text{TP}_t)$ yields the same result. All agents of type $r_i$ are indifferent between experimentation times $\t \in [T_{i-1}, T_i]$.} 
The recursive construction shows that the amount of disclosure under transparency, $\ol H^B_t$, depends on the potential for evidence generation $B(\cdot)$ only through the times $\{T_i\}_{i\le n}$ at which the discount rate of the relevant agent changes from $r_i$ to $r_{i+1}$. 
For the special case of homogeneous agents $n=1$, this implies that the potential for learning $B(\cdot)$ has no impact on welfare. This is in line with the results in \cite{Frick2016}, who study adoption under transparent learning with technology $B(q) = 1-e^{-\l_B q}$. 
 \Cref{prop:OptimalpureBN} with $n=1$ extends their irrelevance result; the learning potential is irrelevant for equilibrium learning beyond exponential generation and even when a designer can control the disclosure of generated evidence.

\section{Pure good-news learning}\label{sec:pureGood}
Consider the opposite extreme where only good evidence can be generated, i.e., $H^B \equiv B \equiv 0$ throughout this section. We show that it is always optimal to delay good news and identify when delay is strictly necessary.

\subsection{Transparent benchmark  under pure good news}

Observing good news does not change the optimal action of agents who are planning to experiment eventually. Therefore, there is no benefit from waiting for such information.\footnote{See also \cite{meyer2025delay}, who focuses on the question of whether waiting for news is valuable in a social-learning game where each player may learn from the opponent's action or outcome. For the setup here, this is exactly the distinction between case 1 and case 2 in \Cref{prop:OptimalPureGN} below.} Experimentation under transparent pure good news occurs at $t=0$:
\begin{lemma}\label{lemma-GN-BM}
    The experimentation path under transparent pure good news is
    \begin{align*}
        q^\text{TP}_t=\begin{cases}
            F_n \quad & \text{ if } v(x(F_n))\geq 0,\\
            \hat{\th}: v(x(\hat{\th}))= 0\quad & \text{ if } v(x(F_n))< 0,
        \end{cases} \qquad \text{ for all } t\ge 0. 
    \end{align*}
In both cases, the expected welfare is $v(x_0)$ for all agents.\footnote{To see that also waiting agents get $v(x_0)$ in the second case, notice that they get $v_G$ with probability $x_0G(\hat{\th})$ and $G(\hat{\th})=v(x_0)/(x_0v_G)$.} 
\end{lemma}
Under pure good news, the belief decreases only gradually. Since no agent wants to invest at beliefs below the myopic threshold, experimentation stops once the belief reaches this threshold. No action-relevant information is generated, and, thus, learning does not improve equilibrium welfare under transparent pure good news.

\subsection{Optimal policy under pure good news}

Delaying good-news disclosure improves welfare when there is potential for the belief to drop below the myopic threshold.
Consider the agent's expected payoff from experimenting at time $\tau$ under good news policy $H^G$:
  \begin{dense}
\begin{align}\label{eq:agent_prob_pureGN}
     \int_0^{\t} e^{-r(\th) t} x_0 v_G \de H^G_t   + e^{-r(\th) \t }
  \Paren{x_0 v_G (1-H^G_\t ) + (1-x_0)v_B }. 
\end{align}
 \end{dense}

First, suppose $v(x(F_n)) \ge 0$. Then we also have $x_0 v_G (1-H^G_\t ) + (1-x_0)v_B \ge 0$ for all $\t$ because $H^G_\t \le G(q_\t)\le G(F_n)$. 
The expression in \eqref{eq:agent_prob_pureGN} is then strictly decreasing in $\t$. All agents optimally invest immediately at $\t =0$, regardless of the disclosure policy. If the expected value is always positive, disclosure cannot overturn the decision to invest, and waiting is fruitless. 
Think of the adoption of clearly superior technologies like messaging apps or USB drives where people switched immediately and later optimism only confirmed a decision that was already best from the start.

Second, suppose that $v(x(F_n)) < 0$.  Then $x_0 v_G (1-H^G_\t ) + (1-x_0)v_B < 0$ whenever $H^G_\t$ is close enough to $G(F_n)$. Now good news is valuable because the lack of disclosure may induce the agent not to invest. 
We show that the disclosure timing of good news plays an important role for welfare whenever a slightly stricter condition holds, $v(x(F_{n-1})) < 0$.

\begin{proposition}\label{prop:OptimalPureGN}
Consider pure good news. 
\begin{enumerate}
\item \label{itm:GN1}If $v(x(F_{n-1})) \geq 0$, welfare is independent of the disclosure policy.
\item \label{itm:GN2} If  $v(x(F_{n-1})) < 0$, the optimal policy satisfies $H^G_t = \begin{cases} 0&\text{if } t< \bar t, \\ G(q_{\bar t-})&\text{if } t\ge \bar t; 
        \end{cases} \; $
        where the time $\bar t>0$ and the amount of experimentation $q_{\bar t-}$ are generically unique and satisfy $v(x(q_{\bar t-}))<0$.
\end{enumerate}
\end{proposition}

The first part of \Cref{prop:OptimalPureGN} shows that the disclosure policy has no effect on welfare when $v(x(F_{n-1})) \geq 0$. The reason disclosure matters only if enough information is generated before the last type starts investing lies in the free order of moves: in equilibrium, all agents of the same type must obtain the same expected payoff as the first of them to invest. If later investors of the same type could secure a higher payoff, the first investor of that type would deviate and wait. Hence, disclosure after the first agent with type $r_n$ has invested cannot benefit any remaining agents.

The second part establishes that whenever $v(x(F_{n-1}))<0$, the designer delays good news to stimulate experimentation beyond the transparent benchmark. In this case, the lack of disclosure changes the optimal action to not investing. To generate incentives for experimentation, the designer must commit to a disclosure delay. Under the optimal policy in part \ref{itm:GN2}, the most impatient agents $\th \in [0,q_{\bar t}]$ invests at time zero, while all others invest if and only if good evidence is revealed at time $\bar t$.

To show that the optimal policy takes this form, we first notice that any optimal policy under the condition in item  \ref{itm:GN2} satisfies $\lim_{\t\to\infty }(x_0 v_G (1-H^G_\t ) + (1-x_0)v_B) <0$. This is necessary for any agent to receive decision-relevant information. The proof in \Cref{proof:OptimalGN} then shows the optimality of a single-time disclosure. Intuitively, patient types benefit more from concentrated disclosure times than impatient types. Hence, one-time disclosure benefits those agents the designer wants to wait and relaxes the incentive compatibility constraint of those agents the designer wants to experiment. Finally, the designer never leaves information on the table: all generated evidence is disclosed at $\bar t$.  

\subsection{Amount of experimentation and comparison with first best}

To finalize the characterization of the optimal policy under pure good news, we present the condition that determines the optimal $\bar t$ in \Cref{prop:OptimalPureGN}. The first step is to compute the minimal delay required to get all agents in $[0,\bar{\theta}]$ to experiment. This delay, denoted by $\bar t(\bar{\theta})$, is determined by the binding incentive constraint of the marginal agent $\bar{\theta}$:
 \begin{align}\label{eq-ICamount}
     v(x_0)=e^{-r(\bar{\theta})\bar{t}(\bar{\theta})}G(\bar{\theta})x_0v_G.
 \end{align}
 We know from \Cref{prop:OptimalPureGN} that $v(x(\bar \theta))<0$, which implies that $G(\bar{\theta})x_0v_G>v(x_0)$.

Using the incentive compatibility constraint \eqref{eq-ICamount}, the designer's problem reduces to
\begin{align}\label{eq:amountOpt}
    \max_{\bar \theta\in[0,F_n] \colon v(x(\bar \theta))<0 } v(x_0) \bar \theta + G(\bar \theta)x_0v_G\int_{\bar \theta}^{F_n}\left( \frac{v(x_0)}{G(\bar \theta)x_0v_G}\right)^{r(\theta)/r(\bar \theta)}d\theta.
\end{align}

As $\bar \theta$ increases, the objective in \eqref{eq:amountOpt} changes in two opposing ways. Each term inside the summation decreases, since the marginal type becomes more patient -- the factor in parentheses is strictly less than one whenever $v(x(\bar \theta))<0$. The terms outside the summation increase in $\bar \theta$, reflecting the greater amount of experimentation and learning. Intuitively, the optimal choice of $\bar \theta$ balances these two forces: the designer selects a larger $\bar \theta$ when there is a big share of very patient agents and when the learning potential is high. Because agents of the same type get the same expected payoff, it is always optimal to have all agents of the same type experiment if some of them do. Thus the optimal amount satisfies $\bar{\theta}\in\{F_1,\dots,F_n\}$.

We compare the optimal incentive-compatible experimentation with the first-best benchmark. Suppose the designer could directly choose investment times on behalf of the agents. Since all delays are inefficient, all investment occurs immediately. The optimal level of experimentation under pure good news, denoted $\theta^{\text{FB}}$, solves
\begin{align*}
\max_{\theta \in [0,F_n]} \; x_0 v_G \left( G(\theta)F_n   +  \bigl(1-G(\theta)\bigr) \theta \right) + (1-x_0) v_B \theta .
\end{align*}

Differentiating this objective and simplifying, we find that the marginal effect of increasing experimentation has the same sign as
\begin{align}  \label{eq:pureGN_FB_FOC}
v(x(\th)) +x(\th) \frac{g(\th)}{1-G(\th)} v_G \Paren{F_n-\th}. 
\end{align}
The first term reflects the expected investment return when agent $\th$ experiments. The second terms captures the learning externality: good news arrives at rate $x(\th) \frac{g(\th)}{1-G(\th)} $, in which case the remaining $F_n-\theta$ agents obtain $v_G$.

A necessary condition for optimality of $\theta^{\text{FB}}<F_n$ is that the first-order condition holds, i.e. \eqref{eq:pureGN_FB_FOC} equals 0.\footnote{This condition is also sufficient and uniquely determines $\theta^{\text{FB}} \in (0,F_n)$ whenever $G$ is concave. For a general $G$, \eqref{eq:pureGN_FB_FOC} may have multiple roots and the condition is only necessary.} This implies that $v(x(\theta^{\text{FB}}))<0$ whenever $\theta^{\text{FB}}<F_n$; the last agents to experiment gets negative expected returns.
Importantly, the first-best condition is independent the distribution of discount rates across agents -- which was the central determinant of optimal second-best experimentation in \eqref{eq:amountOpt}. This contrast yields a surprising result: the second-best may induce more experimentation than the first-best.

\begin{proposition}\label{prop:amountcompNEW3}
Consider pure good news and assume $v(x(F_{n-1})) < 0$ so that good news can have an effect on welfare. The amount of experimentation under any optimal policy 
\begin{itemize}
    \item can be strictly above or below the first-best amount of experimentation,
    \item is strictly above the transparent benchmark.  
\end{itemize} 
\end{proposition}

The social gains from experimentation are always higher in the first best. Over experimentation in the second best arises because the opportunity cost of experimentation is lower than under the first best.
Every agent who does not experiment gets to invest immediately if good news is generated in the first best but must wait until $\bar{t}$ in the second best. This delay reduces the effective cost of experimentation. Proposition \ref{prop:amountcompNEW3} shows that the cost reduction can outweigh the smaller social gains, and the optimal incentive-compatible policy may induce more experimentation than the first-best benchmark.

\section{Delay the good to accelerate the bad}\label{sec:optimal_policy}
Finally, consider joint good and bad news, i.e., $G$ and $B$  strictly increasing. 
We show first that delaying good news is always necessary for optimality as this speeds up the generation of bad news.
We then provide a simple and detail-free optimal policy and illustrate the resulting welfare effects and adoption dynamics. 

\subsection{Necessary conditions under joint news}\label{subsec:NecessaryCdts}

For the formal result statement, define, for a given experimentation process $q$
\begin{align*}
    \bar{n} = \min\{ i \le n \colon q_\infty \le F_i \}, \qquad \text{and for } i < \bar n\colon \; T_i = \inf\{ t \ge 0 \colon q_t > F_{i} \}.
\end{align*}
Hence, $\bar n$ is the last type to experiment and for types $i < \bar{n}$, the first agent of type $i+1$ experiments at time $T_i$. 
Since types $i > \bar{n} $ never experiment,  let $T_{\bar n} = \sup\{ t \ge 0 \colon q_t < q_\infty\}$ be the time of the last experimentation.  $T_{\bar n}$ may be infinite. As a part of the following proposition, we show that all agents of the last experimenting type experiment: $q_{\infty}=F_{\bar{n}}$.

\begin{proposition}\label{prop:main2NEW}
    Suppose $(H^B,H^G, q)$ solve the designer's problem \eqref{eq:MaxProblem_general}. Let $\bar{n}$ be the last type to experiment. Then we must have
    \begin{itemize}
        \item bad news disclosed before next type: ${H}_{t}^B\geq B(F_i)$ for all $t>T_i$ and all $i<\bar{n}$, 
                \item no good news disclosed before last type: $ H^G_t = 0 $ for all $t < T_{\bar{n} -1}$.\, Additionally,  $\bar{n} <n$ only if $v(x(F_{\bar n})) < 0$ and there exists a single disclosure time $\bar t \in (T_{\bar{n} -1}, \infty) $ such that 
      $ H^G_{t} = 0 $  for all $t < \bar t $ and 
        $ H^G_{t} = G(F_{\bar n}) $ 
        for all $t \ge  \bar t $.
    \end{itemize}
\end{proposition}
The first property shows that generated bad evidence must be disclosed before the next type of agents starts experimenting. 
The second property establishes that good evidence must be delayed while agents experiment. The necessity of delays mirrors the case of pure good news in which the absence of good-news generation matters on the action margin (Case \ref{itm:GN2} in \Cref{prop:OptimalPureGN}). 
 Beyond that case, \Cref{prop:main2NEW} shows that delaying good news is always necessary when bad news can also be generated, eliminating the irrelevance result from pure good news (Case \ref{itm:GN1} in \Cref{prop:OptimalPureGN}).

Why does any optimal policy delay good news to speed up bad news and never the reverse? Delaying good news entails a cost: agents invest later in state $G$. The benefit is the accelerated experimentation because (i) without the prospect of future positive disclosures, the payoff from waiting decreases; and (ii) agents are more optimistic absent past disclosures, so the opportunity cost of waiting increases.
When good evidence generated by some type is delayed, the extent to which bad evidence can be accelerated is pinned down by that type's incentive constraint, which equates the marginal cost and benefits of the delay given that type's discount rate.
Then the benefits outweigh the cost for agents who have not adopted yet. 
Recall that bad news is decision relevant on the action margin and good news on the timing margin. 
The remaining, more patient, agents care relatively more about avoiding the wrong action than acting early. Thus, accelerating negative disclosures by delaying positive disclosure strictly benefits them.

To confirm this intuition and prove \Cref{prop:main2NEW}, we solve the designer’s full problem. While the properties of optimal policies with joint news confirm the qualitative insights from the pure-news cases in Sections \ref{sec:pureBad} and \ref{sec:pureGood}, solving the full problem is considerably more demanding. Allowing for joint evidence generation implies that the belief when no news is generated by $q$ experimenters, $x(q)$, need not be monotone and may cross the indifference threshold multiple times for $q \in [0,F_n]$. Consequently, the condition in \Cref{prop:main2NEW} determining whether good evidence must be disclosed at $\bar t$ is expressed in terms of the type $\bar n$ of the last experimenters rather than the  parameter-based case distinction in \Cref{prop:OptimalPureGN}.
More importantly, the dynamic optimization problem with joint evidence involves two control variables subject to interdependent restrictions from the incentive constraints. This motivates our assumption of finitely many types: instead of analyzing an optimal control problem with two controls at each instant and a continuum of (mixed as well as pure state) constraints, we derive necessary conditions via perturbations that yield a finite-dimensional optimization problem for the designer. While a guess-and-verify Lagrangian approach could in principle confirm the optimality of the simple policy in \Cref{prop:main1NEW} below, establishing the necessity of specific features would be daunting. The necessary conditions in \Cref{prop:main2NEW} confirm that any optimal policy exhibits the key features from the pure-news cases -- transparency of bad and delay of good news.

\subsection{A simple optimal policy}
The following is a detail-free optimal policy characterized by a single time $\bar t$ at which good evidence is disclosed if generated by then. All bad evidence is disclosed immediately. 
\begin{proposition}\label{prop:main1NEW}
    There exists a time $\bar t >0$  such that the following policy is optimal
    \begin{itemize}
     \item bad evidence is disclosed immediately:\;    $ H_t^B = B(q_{t-})$\ for all $t$.
     \item good evidence is disclosed at most once: $H_t^G \!= \begin{cases} 0&\text{if } t< \bar t, \\ G(q_{\bar t-})&\text{if } t\ge \bar t. 
        \end{cases} $  
          \end{itemize}
\end{proposition}

\Cref{prop:main1NEW} shows that requiring immediate reporting of breakdowns is consistent with welfare maximization. 
This mirrors regulations governing medical trials with human subjects, where severe adverse effects must be disclosed immediately. For example, the EU regulation governing clinical trials requires ``the immediate cessation of any clinical trial in which there is an unacceptable level of risk.''\footnote{See https://eur-lex.europa.eu/legal-content/EN/TXT/PDF/?uri=CELEX:32014R0536. The regulation refers to the principles from the 1996 Declaration of Helsinki: ``Physicians should cease any investigation if the hazards are found to outweigh the potential benefits.'' 
See https://www.wma.net/wp-content/uploads/2018/07/DoH-Oct1996.pdf.} 
The optimality of immediately reporting bad evidence shows that these safeguarding measures do not conflict with informational interventions targeted at accelerating the initial uptake of experimental treatments. 

Studying financial disclosures, \citet[p. 77]{aboody2000ceo} establish empirically that ``top executives have compensation-related incentives to accelerate the disclosure of bad news and delay announcements of good news.''  The common  practice of compensating CEOs with stock options can create valuable disclosure incentives.\footnote{Other explanations for the finding that managers voluntarily disclose bad news earlier include reputational concerns and litigation risk minimization \citep{skinner1994firms}.}

\paragraph{Welfare effects.}
All agents are better off under the optimal disclosure policy than under transparency. The most impatient agents are indifferent and types $i>1$ benefit. 
\begin{lemma}\label{lm:Pareto}
The disclosure policy in \Cref{prop:main1NEW} results in a Pareto improvement relative to transparent disclosure. 
\end{lemma}
In particular, one can avoid taking a position on interpersonal utility comparisons. There is no tradeoff between potential winners and losers from the intervention, which significantly fosters the acceptance of any proposed measure.

To assess the welfare gain from the designer’s intervention relative to transparency, recall that transparency is optimal under pure bad news (\Cref{prop:OptimalpureBN}) and under pure good news when the absence of positive evidence is not too discouraging (\Cref{prop:OptimalPureGN}, Case \ref{itm:GN1}). Hence, in these cases with optimistic no-news beliefs, the intervention is strictly welfare-improving only when good and bad news are generated jointly (left panel of \Cref{fig:diff_wf_panels_recalc}). 

\Cref{fig:diff_wf_panels_recalc} illustrates the gains for exponential evidence-generation technologies
$B(q) = 1 - e^{-(\l- \D_\l) q}$ and $G(q) = 1 - e^{-(\l+\D_\l) q}$, parametrized by the overall rate $\l>0$ and difference $\D_\l \in [-\l,+\l]$. The hazard rates are $\l_B = 2(\l-\D_\l)$ and $\l_G = 2(\l+\D_\l)$.
The case with optimistic no-news beliefs corresponds to low values of $\l$.\footnote{Specifically, with $\D_\l=\l$, Case \ref{itm:GN1} in Proposition \ref{prop:OptimalPureGN} holds if $2 \l \le \log\Paren{\frac{x_0}{1-x_0} \frac{v_G}{(-v_B)}}/F_{n-1}.$} Here, the welfare gain from optimal disclosure is hump-shaped in $\D_\l$: at both extremes—pure bad news ($\D_\l=-\l$) and pure good news ($\D_\l=+\l$)—transparent disclosure is already welfare-maximizing, while any interior $\D_\l$ yields a strict gain from delaying good news.

Conversely, the right panel shows the case where good-evidence generation is strong enough that no-news beliefs fall below $x^{myop}$, causing experimentation to end inefficiently early under transparency. When $\l$ is large, welfare gains are greatest for high $\D_\l$, as delaying good news sustains a larger amount of experimentation.

With general generation technologies $B$ and $G$, the picture is more nuanced. In broad terms, the largest welfare gains occur when $\frac{1-G(q)}{1-B(q)}$ initially decreases (no news is bad news) before $B$ and $G$ rise sharply for $q > \min{\hat q : v(x(\hat q))=0}$.

\begin{figure}[ht]
  \centering
\begin{tabular}{p{0.45\textwidth}p{0.45\textwidth}}
\vspace{0pt}
\begin{tikzpicture}
  \begin{axis}[
    axis lines = center,
    scale = 0.75,
    xmin = -0.2, xmax = 0.2,
    ymin = 0, ymax = 0.3,
    xtick = {-0.182321557, 0, 0.182321557},
    xticklabels = {$-\l$, $0$, $\l$},
    ytick = {0},
    xlabel = {$\D_\l$},
    ylabel  = {$W^\text{opt} - W^\text{TP}$},
    x label style={at={(current axis.right of origin)},anchor=west},
    y label style={at={(current axis.above origin)},anchor=east},
        extra x ticks={0},
    extra x tick labels={$0$},
    clip=false
  ]
    \addplot [dashed, gray, domain=-0.182321557:0.182321557, samples=2] {0};
    \addplot [very thick] coordinates { {
    (-0.182322,0.000000) (-0.176244,0.010303) (-0.170167,0.020289) (-0.164089,0.029956) (-0.158012,0.039297) (-0.151935,0.048308) (-0.145857,0.056985) (-0.139780,0.065323) (-0.133702,0.073317) (-0.127625,0.080962) (-0.121548,0.088253) (-0.115470,0.095187) (-0.109393,0.101758) (-0.103316,0.107961) (-0.097238,0.113792) (-0.091161,0.119247) (-0.085083,0.124321) (-0.079006,0.129010) (-0.072929,0.133308) (-0.066851,0.137213) (-0.060774,0.140719) (-0.054696,0.143824) (-0.048619,0.146523) (-0.042542,0.148813) (-0.036464,0.150690) (-0.030387,0.152151) (-0.024310,0.153193) (-0.018232,0.153815) (-0.012155,0.154014) (-0.006077,0.153788) (0.000000,0.152923) (0.006077,0.152059) (0.012155,0.150554) (0.018232,0.148625) (0.024310,0.146271) (0.030387,0.143496) (0.036464,0.140303) (0.042542,0.136697) (0.048619,0.132684) (0.054696,0.128271) (0.060774,0.123468) (0.066851,0.118286) (0.072929,0.112740) (0.079006,0.106844) (0.085083,0.100619) (0.091161,0.094088) (0.097238,0.087277) (0.103316,0.080219) (0.109393,0.072951) (0.115470,0.065517) (0.121548,0.057971) (0.127625,0.050374) (0.133702,0.042801) (0.139780,0.035340) (0.145857,0.028101) (0.151935,0.021215) (0.158012,0.014847) (0.164089,0.009204) (0.170167,0.004562) (0.176244,0.001298) (0.182322,0.000000)
    } };
  \end{axis}
\end{tikzpicture}
&
\vspace{0pt}
\begin{tikzpicture}
  \begin{axis}[
    axis lines = center,
    scale = 0.75,
    xmin = -0.8, xmax = 0.8,
    ymin = 0, ymax = 38.6,
    xtick = {-0.75, 0, 0.182321557, 0.75},
    xticklabels = {$-\lambda$, $0$, $\Delta^{\dagger}$, $\lambda$},
    ytick = {0},
    xlabel = {$\D_\l$},
    ylabel = {$W^\text{opt} - W^\text{TP}$},
    x label style={at={(current axis.right of origin)},anchor=west},
    y label style={at={(current axis.above origin)},anchor=east},
       extra x ticks={0},
    extra x tick labels={$0$},
    clip=false
  ]
    \addplot [dashed, gray, domain=-0.75:0.75, samples=2] {0};
    \addplot [very thick] coordinates { {(-0.750000,0.000000) (-0.731487,0.089827) (-0.712975,0.179869) (-0.694462,0.270120) (-0.675949,0.360576) (-0.657437,0.451233) (-0.638924,0.542088) (-0.620411,0.633138) (-0.601899,0.724380) (-0.583386,0.815813) (-0.564873,0.907436) (-0.546361,0.999249) (-0.527848,1.091252) (-0.509335,1.183450) (-0.490823,1.275845) (-0.472310,1.368443) (-0.453797,1.461252) (-0.435285,1.554284) (-0.416772,1.647549) (-0.398259,1.741066) (-0.379747,1.834853) (-0.361234,1.928936) (-0.342722,2.023343) (-0.324209,2.118110) (-0.305696,2.213280) (-0.287184,2.308903) (-0.268671,2.405039) (-0.250158,2.501759) (-0.231646,2.599149) (-0.213133,2.697310) (-0.194620,2.796362) (-0.176108,2.896449) (-0.157595,2.997740) (-0.139082,3.100442) (-0.120570,3.204802) (-0.102057,3.311116) (-0.083544,3.419748) (-0.065032,3.531139) (-0.046519,3.645836) (-0.028006,3.764516) (-0.009494,3.888032) (0.009019,4.017467) (0.027532,4.154221) (0.046044,4.300135) (0.064557,4.457686) (0.083070,4.630300) (0.101582,4.822890) (0.120095,5.042859) (0.138608,5.302178) (0.157120,5.622532) (0.175633,6.053504) (0.194146,7.379851) (0.212658,9.126114) (0.231171,10.824731) (0.249684,12.408821) (0.268196,13.920120) (0.286709,15.363859) (0.305222,16.744866) (0.323734,18.067316) (0.342247,19.334539) (0.360759,20.548878) (0.379272,21.755650) (0.397785,22.891378) (0.416297,23.986159) (0.434810,25.042039) (0.453323,26.060732) (0.471835,27.043448) (0.490348,27.990545) (0.508861,28.900939) (0.527373,29.847191) (0.545886,30.720330) (0.564399,31.566279) (0.582911,32.383909) (0.601424,33.169470) (0.619937,33.985807) (0.638449,34.743936) (0.656962,35.474866) (0.675475,36.159260) (0.693987,36.914440) (0.712500,37.67)} };
  \end{axis}
\end{tikzpicture}
\end{tabular}
\caption{\linespread{1.0}\selectfont \small Welfare difference between optimal policy and transparency with $n=2$ and exponential generation technology. Left panel: overall  evidence potential $\l$ low,  so that even when $\D_\l = \l$, the no-news belief remains above the myopic indifference threshold. Right panel: $\l$ large so that for and $\D_\l > \D^\dagger$, the no-news belief eventually crosses   the myopic threshold.}
\label{fig:diff_wf_panels_recalc}
\end{figure}

\paragraph{Adoption dynamics.} Finally, we discuss the investment and belief dynamics under the policy in \Cref{prop:main1NEW}. 
Figures \ref{fig:DynamicsBN} and \ref{fig:DynamicsGN} illustrate the amount of experimentation and the belief, both under the optimal policy and under transparency. We again depict exponential evidence technologies with hazard rates $\l_B$ and $\l_G$. \Cref{fig:DynamicsBN} considers the case where no news is good news ($\l_B>\l_G$); in \Cref{fig:DynamicsGN}, no news is bad news ($\l_B <\l_G$). Beliefs are depicted in log-likelihood ratios, $\ell_t=ln(x_t/(1-x_t))$.
As \Cref{fig:DynamicsBN} has $\l_B>\l_G$, the no-disclosure belief under transparency increases. This increase is even faster under the optimal policy because good news are censored and experimentation is faster. 

\begin{figure}[ht]
  \centering
\begin{tabular}{p{0.45\textwidth}p{0.45\textwidth}}
\vspace{0pt}  
\begin{tikzpicture}
	 \begin{axis}[
                        axis lines = center,
                        scale = 0.75,
                        xtick = {0.001,.192,.284},
                        xticklabels = {$0$,${T_1}$,${T_2}$},
                        ytick = {0.001,.5,1},
                        yticklabels = {$0$,$F_1$,$F_2$},
                        xmin = 0,
                        xmax = .5,
                        ymin = 0,
                        ymax = 1.2, 
                        xlabel = {$t$},
                        ylabel = {
                        },
                        x label style={at={(current axis.right of origin)},anchor=west},    
                        y label style={at={(current axis.above origin)},anchor=south},    
                    ]
                    
                 \addplot [dashed, gray, domain=0:.45, samples=2]{1} ;   
                    \addplot [dashed, gray, domain=0:.192, samples=2]{.5} ;

                    \addplot [ gray, domain=0:.245, samples=50]{1/6*(x*3-3*ln(2*exp(3*x/3)-exp(3*x))}
                ;
             
                    \addplot [ gray, domain=.245:.445,samples=50]{ 1/6*(.735+x-0.245-3*ln(2*exp((.735+x-0.245)/3)-exp(.735+x-0.245)))}
                         node[below,pos=.05] {$q^\text{TP}$}; 
 
  		 \addplot [gray, domain=.445:.5,samples=2]{1}; 	

                              \draw[-,  blue, dashed, thick] (axis cs:0.048,0.056) to  (axis cs:0.495,0.056);
            \fill[blue] (axis cs:0.1625,.056) circle (0.5pt);
                          \fill[blue] (axis cs:0.175,.056) circle (0.5pt);
                 \fill[blue] (axis cs:0.1875,.056) circle (0.5pt);


                              \draw[-,  blue, dashed, thick] (axis cs:0.144,0.26) to  (axis cs:0.49,0.26);

                              \draw[-,  blue, dashed, thick] (axis cs:0.238,0.662) to  (axis cs:0.5,0.662);

                              \draw[-,  blue, dashed, thick] (axis cs:0.275,0.9) to  (axis cs:0.5,0.9);
                              
                              \addplot [very thick, domain=0:.192, samples=50]{1/3*ln(1/(2-exp(3*x)))};

                       \addplot [very thick, domain=.192:.284,samples=50]{ 1/3*ln(1/(2-exp(0.576 + x- .192)))}
                       node[above,pos=.25] {$q^*$}
                       ; 
                    \addplot [very thick, domain=.2844:.5,samples=2]{1}; 
                    \end{axis}
		\end{tikzpicture} & \vspace{0pt} %
\begin{tikzpicture}
	 \begin{axis}[
                        axis lines = center,
                        scale = 0.75,
                        xtick = {.001,.192,.284},
                        xticklabels = {0,${T_1}$,${T_2}$},
                        ytick = {.225, 0.35},
                        yticklabels = {$\ell^\text{myop}$, $\ell_0$},
                        axis y discontinuity=parallel,
                        xmin = 0,
                        xmax = .5,
                        ymin = 0,
                        ymax = 1.2, 
                        xlabel = {$t$},
                        ylabel = {
                        },
                        x label style={at={(current axis.right of origin)},anchor=west},    
                        y label style={at={(current axis.above origin)},anchor=south},    
                    ]
                    
                    \addplot [dashed, gray, domain=0:.5, samples=2]{.225} ;

                    \addplot [ gray, domain=0:.245, samples=50]{0.35+1/10*(x*3-3*ln(2*exp(3*x/3)-exp(3*x))}
                ;
             
                    \addplot [ gray, domain=.245:.445,samples=20]{ 0.35+1/10*(.735+x-0.245-3*ln(2*exp((.735+x-0.245)/3)-exp(.735+x-0.245)))}
                         node[below,pos=.6] { $\ell^{\text{TP}}$}; 
                         \addplot [ gray, domain=.445:.52,samples=50]{ 0.95};
 

                    \draw[->,  blue, >=stealth, dotted, thick] (axis cs:0.048,0.392) [bend left] to  (axis cs:0.048,0.05);
                 \draw[->,  blue, >=stealth, dotted, thick] (axis cs:.144,0.544) [bend left] to  (axis cs:.144,0.05);
                                           
                       \draw[->,  blue, >=stealth, dotted, thick] (axis cs:0.238,0.846) [bend left] to  (axis cs:0.238,0.05);
                        \draw[->,  blue, >=stealth, dotted, thick] (axis cs:0.275,1.025) [bend left] to  (axis cs:0.275,0.05);

                              \addplot [very thick, domain=0:.192, samples=50]{0.35+1/4*ln(1/(2-exp(3*x)))};
                       \addplot [very thick, domain=.192:.284,samples=50]{ 0.35+1/4*ln(1/(2-exp(0.576 + x- .192)))}
                       node[above,pos=.25] {$\ell^*$}
                       ; 
                    \addplot [very thick, domain=.2844:.5,samples=2]{1.1}; 
                    \end{axis}
		\end{tikzpicture}
\end{tabular}%
\caption{\linespread{1.0}\selectfont \small Experimentation $q_t$ and no-disclosure belief $\ell_t=ln(x_t/(1-x_t))$ when $\l_B>\l_G$. At all times prior to $T_2$, negative evidence is disclosed at a positive rate. If negative evidence is disclosed, there is no further investment (left plot, dashed blue lines) and the belief jumps to 0, i.e., the log-likelihood ratio jumps to $-\infty$ (right plot, blue arrows). }
\label{fig:DynamicsBN}
\end{figure}

\begin{figure}[ht]
  \centering
\begin{tabular}{p{0.45\textwidth}p{0.45\textwidth}}
\vspace{0pt}  
\begin{tikzpicture}
	 \begin{axis}[
                        axis lines = center,
                        scale = 0.75,
                        xtick = {0.001,.1275,.3},
                        xticklabels = {$0$,${T_1}$,${\bar t}$},
                        ytick = {0.001,.7,1},
                        yticklabels = {$0$,$F_1$, $F_2$},
                        xmin = 0,
                        xmax = .5,
                        ymin = 0,
                        ymax = 1.2, 
                        xlabel = {$t$},
                        ylabel = {
                        },
                        x label style={at={(current axis.right of origin)},anchor=west},    
                        y label style={at={(current axis.above origin)},anchor=south},    
                    ]
                    
                \addplot [dashed, gray, domain=0:.47, samples=2]{1} ;   
                    \addplot [dashed, gray, domain=0:.47, samples=2]{.7} ;   

 \addplot [ gray, domain=0:.47, samples=50]{10*x/(20*x+3)}node[below,pos=.65] { $q^{\text{TP}}$}
                ;

                ;
             
 

                              \draw[-,  blue, dashed, thick] (axis cs:0.09,0.412) to  (axis cs:0.47,0.412);


                              \draw[-,  blue, dashed, thick] (axis cs:0.12,0.632) to  (axis cs:0.465,0.632);


                       ; 
  
                 \draw[->, thick, orange, >=stealth, dashed] (axis cs:0.3,0.7) to (axis cs:0.3,.985);
               \fill[orange] (axis cs:0.3,1) circle (1.5pt);
                 
                 
                 \draw[thick, orange] (axis cs:0.3,1) -- (axis cs:0.47,1);

                      \addplot [very thick, domain=0:.1275, samples=50]{10/9*ln(1/(2-exp(3*x)))}node[left,pos=.75] {$q^*$};

  \addplot [very thick, domain=.1275:.47,samples=2]{0.7}; 
                  
              \end{axis}
		\end{tikzpicture} & \vspace{0pt} %
\begin{tikzpicture}
	 \begin{axis}[
                        axis lines = center,
                        scale = 0.75,
                        xtick = {0.001,.1275,.3},
                        xticklabels = {$0$,${T_1}$,${\bar t}$},
                        ytick = {.35, 0.5},
                        yticklabels = {$\ell^\text{myop}$, $\ell_0$},
                        axis y discontinuity=parallel,
                        xmin = 0,
                        xmax = .5,
                        ymin = 0,
                        ymax = 1.2, 
                        xlabel = {$t$},
                        ylabel = {
                        },
                        x label style={at={(current axis.right of origin)},anchor=west},    
                        y label style={at={(current axis.above origin)},anchor=south},    
                    ]
                    
                    \addplot [dashed, gray, domain=0:.5, samples=2]{.35} ;

                    \addplot [ gray, domain=0:.47, samples=50]{0.35+1.5/(150*x+10)}node[above,pos=.5] { $\ell^{\text{TP}}$}
                ;
             
 

                    \draw[->,  blue, >=stealth, dotted, thick] (axis cs:0.06,0.555) [bend left] to  (axis cs:0.06,0.05);
                       \draw[->,  blue, >=stealth, dotted, thick] (axis cs:0.12,0.642) [bend left] to  (axis cs:0.12,0.05);

                 \draw[->, thick, orange, >=stealth, dashed] (axis cs:0.3,0.67) to  (axis cs:0.3,1.15);

        \addplot [very thick, domain=0:.1275, samples=50]{0.5+1/4*ln(1/(2-exp(3*x)))}node[above,pos=.5] {$\ell^*$};
        
                      \addplot [very thick, domain=.1275:.3,samples=2]{.66};
                    \addplot [very thick, domain=.3:.47,samples=2]{.2}; 
                   \draw[thick, dashed] (.3,.66) -- (.3,.2);

                    \end{axis}
		\end{tikzpicture}
\end{tabular}%

\caption{\linespread{1.0}\selectfont \small Experimentation and no-disclosure belief when $\l_B<\l_G$ and $x(F_1)<x^\text{myop}$.  
 Any time prior to $ T_1$, negative evidence is disclosed at a positive \textit{rate}. Between $ T_1$ and $\bar t$, nobody invests and no evidence is disclosed. At time $\bar t$, positive evidence is disclosed with positive \textit{probability}. If positive evidence is disclosed, the belief jumps to 1 and all agents invest (orange arrow). If no positive evidence is disclosed, the no-disclosure belief jumps below the indifference threshold and no further agents invest.}
\label{fig:DynamicsGN}
\end{figure}

\Cref{fig:DynamicsGN} has $\l_B< \l_G$, so the no-disclosure belief under transparency would decrease. 
However, since the optimal policy delays all good news until time $\bar t $, the no-disclosure belief is increasing. 
\Cref{fig:DynamicsGN} depicts an example with $(\l_G -\l_B )F_1$ large enough so that $x(F_1)< x^\text{myop}$. 
This means that the posterior belief drops below the myopic threshold once we learn that the investment by all agents with type $r_1$ failed to generate evidence. Then, under transparency, not all high-type agents would experiment: the total amount of experimentation under transparency approaches the level at which the no-disclosure belief equals the myopic indifference threshold. 
In contrast, the optimal policy delays good news so that all type-$r_1$ agents experiment. 
Eventually, the designer discloses all generated good evidence at time $\bar t$, at which the remaining agents invest if good news is disclosed. 
To deter type-$r_1$ agents from waiting, the disclosure time $\bar t$ must be sufficiently late. 
Since there is a strictly positive probability of disclosure at time $\bar t$, there is a period $({T}_1,\bar t )$ during which no agent experiments and no information is disclosed.

Related to the over-experimentation documented in \Cref{prop:amountcompNEW3} for pure good news, \Cref{fig:DynamicsGN} reveals an additional benefit of over-experimentation under joint news. Agents become more optimistic as the optimal policy commits do disclose bad news. Thus when more agents experiment, the last experimenter is more optimistic. This makes him less tempted to wait and reduces the necessary delay $\bar t - T_{\bar n}$. 

\section{Discussion} \label{sec:discussion}

\subsection{Homogenous agents and alternative design objectives}

Since the payoff of the first agent to invest is fixed, the disclosure policy affects welfare only if agents are heterogeneous. Disclosure can be relevant with  homogeneous agents under different design objectives. Online Appendix \ref{sec-example} contains an example showing that our policy maximizes the speed at which all available evidence is disclosed.
The transparency result \Cref{prop:OptimalpureBN} is independent of the design objective and carries over.

Studying heterogeneous agents, we focus on heterogeneity in discount rates for three reasons. 
First, the fundamental inefficiency in the transparent benchmark and the second-best policy consists of delays, and we analyze optimal disclosure policies when individuals differ precisely in how much delays affect them. 
Second, discounting heterogeneity has the analytical advantage that the myopic indifference threshold is the same for everyone, which implies that whether each type of evidence is decision relevant or not is the same for all agents. 
Third, discount rates can capture differences in the rate at which the investment opportunity becomes unavailable or obsolete for different agents, e.g., a vaccine becomes useless when the agent gets infected by a virus before being vaccinated.
Other forms of heterogeneity, or a combination of them, are also natural. We present the two most obvious alternatives and discuss how our results extend to them. 

\vspace{0.18cm} 

\noindent
\textbf{Payoff heterogeneity.} 
 Consider a model otherwise identical to the main model but suppose that the types $1,\dots,n$ differ in their investment payoffs such that  $v_{\omega}^{i}\geq v_{\omega}^{i+1}$ for all $i$ and $\omega$. This leads to investment dynamics similar to the main model where types with lower indices always invest earlier.

\vspace{0.18cm} 

\noindent
\textbf{Different arrival times.} 
 For another form of heterogeneity consider (identical) agents who arrive in separate cohorts. To fix ideas, suppose that type $i$ specifies  the arrival time in $\{t_1,\dots,t_n\}$ with $t_i < t_{i+1}$. Now, if $t_{i+1} - t_{i}$ is short enough, not all cohort $i$ agents invest before cohort $i+1$ arrives under transparency.\footnote{In the other case, $t_{i+1} - t_{i}$ is sufficiently large for all $i$, there is no scope for the disclosure policy to affect welfare.} 

The bad-news result in \Cref{prop:OptimalpureBN} extends to both alternatives above. The argument in the proof shows that the agents' incentive constraints cannot be satisfied if there is more information than under transparency \textit{at any future time}. The argument involves no comparison across types and, carries when agents differ in other dimensions. 

\subsection{Manipulating evidence}

In our setup, the designer cannot manipulate or fabricate evidence but decides when to disclose hard evidence. In comparison to more permissive information design tools where the sender can send any messages based on what she has learned, we believe that choosing the timing at which gathered evidence is released as is requires less commitment power. 
Furthermore, considering a hard evidence model without freedom over what messages can be sent helps to focus on the question of \textit{when} information should be revealed. However, it may be interesting to ask what happens if the designer can manipulate evidence.

Studying welfare maximization in the classical Bayesian-persuasion approach would lead to a trivial result: if the designer can perfectly condition the messages on the true state of the world, there is no reason to hide any information. 
Even if the designer needs to learn the state through endogenous experimentation,  full control over the information environment, including the ability to send private messages, would allow her to implement the first best. 
The tradeoff between generation and disclosure does not arise if the designer can exclude agents from future disclosures.

The case where the designer learns the state endogenously and flexibly designs public messages is closer to our problem. \cite{chen2024optimal} study this design problem under pure good and pure bad news respectively and find that faking bad news can be optimal under a finite investment horizon, when there is not enough time for everyone to invest in the transparent equilibrium. 
We conjecture that faking bad news does not increase welfare in our pure bad news case with infinite horizon. However, a full analysis of optimal public evidence manipulation is beyond the scope of the present paper. Under joint news, the space of feasible belief processes alone is hard to track. 

\subsection{Concluding remarks}

We study the dynamic tradeoff between information generation and disclosure. How should data be disclosed to the public when the data-generating process is endogenously determined by agents who base their choices on past and anticipated future disclosures?

The main takeaway is that information revealing bad quality should be disclosed promptly, while good news should be collected for a longer time and disclosed with delay. This finding resonates with regulation demanding that fatal breakdowns be reported immediately. Conversely, it may be beneficial to keep experimenting before releasing a potential breakthrough. If the public expects to get to know about breakthroughs fast, they may choose to wait until one is disclosed and become pessimistic in its absence.

More broadly, the fact that the data-generating process is shaped by expectations about how data will be disclosed or used in the future appears relevant in many economic settings.
We believe that analyzing dynamic design problems -- of information, allocations, or both -- through this lens offers a promising avenue for future research.


\newpage

\appendix

\section{Main proofs and additional derivations}

\subsection{Proof of \Cref{prop:OptimalpureBN}}\label{proof:OptimalBN}

Let $x_t$ and $q_t$ denote the belief and the stock of experimenters under an arbitrary implementable policy $H^B$, and similarly $x_t^\text{TP}$ and $q_t^\text{TP}$ under transparent breakdowns.
Suppose, contrary to the claim, that $H_t^B>B(q_t^\text{TP})$ for some $t$. Let $\hat{t}:=\inf\{\tau\in\mathbb{R}_+:H_{\tau}^B>B(q_{\tau}^\text{TP})\}$.  Notice that $B(q_t^\text{TP})$ is continuous; in particular $q_t^\text{TP}=q_{t-}^\text{TP}$.
As a preliminary step, note that ${q}_{t}-{q}_{t-}>0$ only if $H_{t-}^B=H_{t+}^B$ because waiting at $t$ would be strictly optimal for all types if bad news were disclosed with strictly positive probability. Hence, we have either $q_{\hat{t}-}>q_{\hat{t}}^\text{TP}$ or $q_{\hat{t}-}=q_{\hat{t}}^\text{TP}$ and $q_{\hat{t}+\e}>q_{\hat{t}+\e}^\text{TP}$ for $\e>0$ small enough.

The ordering of $q$ and $q^\text{TP}$ implies that there exist times $t'<t''$ with $t' \le\hat{t}\le t''$ such that some agents with some type $r$ experiment at $t'$ under $H^B$ but experiment at $t''$ under transparency. Furthermore, $t'$  satisfies $H_{t'}^B\leq B(q_{t'}^\text{TP})$, and $t''$  satisfies $H_{t''}^B> B(q_{t''}^\text{TP})$.\footnote{We can choose $t''$ such that the latter holds because agents experiment continuously under transparency so the ordering of disclosed information persists for some interval close to $\hat{t}$.}

Since experimenting at $t''$ is optimal under transparent breakdowns, we have
   \begin{align}
   x^\text{TP}_{t'}v_G+(1-x^\text{TP}_{t'})v_B \leq  e^{-r(t''-t')}\left(x^\text{TP}_{t'} v_G+(1-x^\text{TP}_{t'})\frac{1-B(q^\text{TP}_{t''})}{1-B(q^\text{TP}_{t'})}v_B\right).\label{eq:ProofProp1Waiting}
\end{align}
Since experimenting at $t'$ is optimal under the alternative policy $H^B$, we have
\begin{align}
    x_{t'}v_G+(1-x_{t'})v_B \geq 
    e^{-r(t''-t')}\left(x_{t'} v_G+(1-x_{t'})\frac{1-H^B_{t''}}{1-H^B_{t'}}v_B\right).\label{eq:ProofProp1Experimenting}
\end{align}
Given that $x_{t'}^\text{TP}\geq x_{t'}$ (because $ B(q_{t'}^\text{TP}) \ge H_{t'}^{B}$)
and  
$(H_{t''}^B-H_{t'}^{B})/(1-H_{t'}^{B})> (B(q_{t''}^\text{TP})-B(q^\text{TP}_{t'}))/(1-B(q^\text{TP}_{t'}))$ (because, additionally, $ B(q_{t''}^\text{TP}) < H_{t''}^{B}$), \eqref{eq:ProofProp1Waiting} and \eqref{eq:ProofProp1Experimenting} cannot hold simultaneously: 
divide both sides of \eqref{eq:ProofProp1Experimenting} by $x_{t'}$ and subtract it from \eqref{eq:ProofProp1Waiting} divided by $x^\text{TP}_{t'}$
to get
\begin{align*}
    (-v_B) \Paren{\frac{1-x_{t'}}{x_{t'}}- \frac{1-x^\text{TP}_{t'}}{x^\text{TP}_{t'}}} \le     (-v_B)e^{-r(t''-t')}\Paren{ \frac{1-x_{t'}}{x_{t'}} \frac{1-H^B_{t''}}{1-H^B_{t'}} - \frac{1-x^\text{TP}_{t'}}{x^\text{TP}_{t'}}\frac{1-B(q^\text{TP}_{t''})}{1-B(q^\text{TP}_{t'})}},
\end{align*}
 which yields a contradiction. 
 \qed

\subsection{Proof of \Cref{lemma-GN-BM}}
 (i) Let $x(F_n)\geq x^\text{myop}$. Then $x(q)\geq x^\text{myop}$ for all $q\in[0,F_n]$ as $x(q)$ is decreasing under pure good news. This implies that all agents invest both with and without good news. Because of discounting, they are better off investing at $t=0$. 
(ii) Let $x(F_n)< x^\text{myop}$ and let $\hat{\theta}$ be as defined in the statement. Clearly, no agent experiments if $q^\text{TP}_t>\hat{\theta}$ as then $v(x_t)<0$. Similarly, if we had $q^\text{TP}_t<\hat{\theta}$ at any $t>0$, agents would have a strict preference to invest. Hence, $q^\text{TP}_t=\hat{\theta}$ for all $t$. 
\qed

\subsection{Proof of \Cref{prop:OptimalPureGN}}\label{proof:OptimalGN}

Suppose first that $v(x(F_{n-1})) \geq 0$. Define $\hat \th: v(x(\hat \th)) = 0$. If $H^G_t\leq G(\hat \th)$ for all $t$, then all agents invest both with and without good news, and hence their welfare is independent of the disclosure policy. Suppose $H^G_t> G(\hat \th)$ for some $t$, and define $\tau:=\sup\{t:H^G_t\leq G(\hat \th)\}$. As  $\hat \th>F_{n-1}$, some type-$n$ agents (and all agents with $i<n$) must experiment at time $\tau$ or before, which means that they invest regardless of news. As all agents with the same type get the same expected payoff in any equilibrium, it means that all agents get the same expected welfare as if no information were ever disclosed.

Second, let $v(x(F_{n-1})) < 0$. 
We show that under pure good news $\t(\th) < \infty $ implies $\t(\th)=0$ for all $\th$. If agent $\th$ is willing to experiment, then he invests at time 0. 
To see this, note that $\tau(\th) < \infty$ only if $v(x_{\tau(\th)}) \ge 0$, or, equivalently, $x_0 v_G (1-H^G_{\tau (\th)}) +(1-x_0) v_B \ge 0$.
This implies that the expected payoff from experimenting at $\t(\th)$ is smaller than $v(x_0)$:
\begin{align*}
&\int_0^{\t(\th)} e^{-r(\th) t} x_0 v_G \de H^G_t + e^{-r(\th)\t(\th)}\Paren{ x_0 v_G (1-H^G_{\t(\th)}) + (1-x_0)v_B}\leq  x_0 v_G H^G_{\tau (\th)}
\\
& 
  +  e^{-r(\th)\t(\th)}\Paren{ x_0 v_G (1-H^G_{\t(\th)}) + (1-x_0)v_B}
= v(x_0) -(1-e^{-r(\th)\t(\th)})v(x_{\tau(\th)}).
\end{align*}
 The inequality is strict unless $\lim_{t\downarrow 0}H^G_{t}=H^G_{\tau (\th)}=v(x_{0})/(x_0v_G)$. If the condition holds, and if in addition  $\lim_{t\uparrow \infty}H^G_{t}=v(x_{0})/(x_0v_G)$, the agents are indifferent between all experimentation times. In that case, everyone gets the same expected payoff as if $H^G_{t}=0$ for all $t$, which is the lower bound an incentive compatible policy can yield to any agent. We verify that the optimal policy we construct gives a strictly higher payoff to some agents.
Hence, $\t(\th) \in \{0,\infty\}$ for all $\th \in [0,F_n]$ in any optimal policy. 

Let $\bar n = \max\{ i \le n \colon \exists \th \text{ with } (r(\th) = r_i  \cap \t(\th)=0) \}$, be the most patient type who experiments at time 0. 
Since waiting is costlier for all more impatient types, a disclosure policy with a fixed $\bar n$ is incentive compatible if and only if $
v(x_0) \ge x_0 v_G \int_0^{\infty} e^{-r_{\bar n} t}  \de H^G_t. 
$
The total welfare under pure good news satisfies 
$
F_{\bar n} v(x_0)  + x_0 v_G \sum_{i > \bar n} f_i \int_0^{\infty} e^{-r_i  t}  \de H^G_t. 
$

Next, we show that disclosing good news at most once is optimal. Take an implementable disclosure policy $H_t^G$, and let $H^G_\infty = \lim_{t\to\infty}H_t^G$.
We construct an alternative policy $\tilde{H}_t^G$ that discloses all information at time $\bar t$: $\tilde{H}_t^G=0$ if $t<\bar t$ and $\tilde{H}_t^G={H}_\infty^G$ for all $t\geq\bar t$. Time $\bar t$ is chosen such that the agents of type $\bar n$ get the same value of waiting under the original and alternative policy. 
 Thus, $\bar t$ satisfies
    \begin{align}\label{eq:proof-GN-bart}
&e^{-r_{\bar n}  \bar t } x_{ 0}v_G H^G_\infty  =
x_{0} v_G\int_{ 0}^{ \infty} e^{-r_{\bar n} t}  \de H_t^G.
\end{align}
As type $\bar n$ is willing to experiment at $t=0$ under the original policy, he is also willing to do so under the alternative policy; this also implies that more impatient agents are not induced to wait longer by the modification, ensuring feasibility of $\tilde H^G$.   
With $\bar t$ defined by \eqref{eq:proof-GN-bart}, the more patient types $k > \bar n$ are strictly better off under the alternative policy than the original policy because 
\begin{align}\label{eq:lm1-GN-improvement}
    e^{-r_k \bar t}  > \frac{ \int_{0}^\infty e^{- r_k t} \de H_t^G}{H^G_\infty}, \quad \text{ for all } k > \bar n.
\end{align}
To see this,  \eqref{eq:proof-GN-bart} implies that  $ e^{- r_k \bar t}$ is equal to
$ \Big (  \frac{ \int_{0}^\infty e^{- r_{\bar n} t} \de H_t^G}{H^G_\infty}\Big )^{r_k / r_{\bar n}}$.
Since $r_k/r_{\bar n } < 1$, Jensen's inequality implies that this is greater than the right-hand side of \eqref{eq:lm1-GN-improvement}; more patient agents are strictly better off. 

Finally, it is optimal to disclose all generated evidence, i.e. $H_{\bar t}^G=G(q_{\bar t})$ when $\bar n <n$. Suppose not, and let the amount of good evidence disclosed at $\bar t$ be ${H}^{G}_\infty < G(q_{\bar t})$. Then, The same steps as above, show that types $k>\bar n$ are better off if the designer discloses all information $G(q_{\bar t})$ at $\bar t+\D$, where $\D$ is chosen such that type $\bar n$ remains indifferent: $e^{- r_{\bar n} \D}=\frac{{H}^{G}_\infty}{G(q_{\bar t})}$.
 Thus, whenever $\bar n < n$, all generated information is disclosed. 
\qed

 \subsection{First-best benchmark}\label{sec:FBdetailed}
 
 As explained in the main text, the first best features no delays. 
Yet, to learn from previous adopters, agents should still experiment sequentially so that later agents stop experimenting if an earlier adopter generated bad news.
  As a result, the strategy space introduced in our main model cannot directly capture the first best strategy profile when there is bad evidence. 
To overcome this, we impose an exogenous bound on the experimentation rate such that $\dot q_t \le \rho$ and consider the limit as $\rho  \to \infty$.\footnote{Alternatively, we could extend the strategy space to allow agents to act sequentially at each instant, adding a \textit{virtual time} dimension.}
 
Fix $\rho >0$ and consider the total expected welfare if the agents $[0,\bar \th]$ experiment.\footnote{In the limit as $\rho \to \infty$, the order of agents becomes irrelevant as every agent's discount factor converges to 1, irrespective of type $r$. For $\rho<\infty$, having impatient agents experiment earlier is optimal, so we let agents experiment in increasing order of $\th$ also in the first best benchmark.} 
We have $\t(\th) = \th/\rho$ for all $\th \le \bar \th$ and $\tau(\th)=\infty$ for all $\th>\bar \th$ since the latter types only invest in case of good news. Further, $\de H^G_s = g(q_s)\dot q_s \de s$ with   $\dot q_s = \rho $ and $q_s = \rho s$. 
\begin{align*}
 &\int_{0}^{\bar \th } \Big [ \int_0^{ \th /\rho } e^{-r(\th) s} x_0 v_G g(\rho \, s) \rho \de s  + e^{-r(\th) \th/\rho}\Paren{x_0 (1-G(\th))  v_G + (1-x_0)(1-B(\th) )v_B}  \Big ]\de \th
 \\
 + & \int_{\bar \th}^{F_n} \Big[ \int_0^{ \bar \th /\rho } e^{-r(\th) s} x_0 v_G g(\rho \, s) \rho \de s  \Big] \de \th 
\end{align*}

Taking the derivative wrt $\bar \th$ and
 dividing by $ e^{-r(\bar \th) \bar{\th} /\rho} (x_0 (1-G(\bar \th))+ (1-x_0)(1-B(\bar \th) ))$, the marginal effect of having agent $\bar \th$ experiment has the same sign as $
v(x(\bar \th)) + x(\bar \th)  g(\bar \th )/(1-G(\th))v_G(F_n-\bar \th  )$.
This expression is independent of $\rho$ and equals the condition in the main text, with the difference that $x(\th)$ incorporates both good and bad evidence. 
If $v(x(\th)) \ge 0$ for all $\th$ -- which is always the case when $B \ge G$ -- then all agents should experiment. Otherwise, if $v(F_n) <0$, then the first best amount of experimentation is strictly below $F_n$ because the informational benefit vanishes as $\bar \th$ approaches $F_n$.

\subsection{Proof of \Cref{prop:amountcompNEW3}}\label{Aproof-prop:amountcompNEW3}
The second part, that the optimal policy implements weakly more experimentation than the transparency benchmark, follows directly from \Cref{prop:OptimalPureGN} as $x(q_{\bar t})<x^{myop}$ if $x(F_{n-1}) < x^\text{myop}$, while the transparency benchmark has $\lim_{t\to\infty}x_t=x^{myop}$ by \Cref{lemma-GN-BM}.

To prove the first part, we first show that the optimal amount of experimentation in the second best $\bar{\theta}$ is such that all agents with the same type take the same action: $\bar{\theta}\in\{F_1,\dots,F_n\}$. Suppose the policy takes the form of \Cref{prop:OptimalPureGN} but that $q_{\bar t}\in (F_i,F_{i+1})$ for some $i\in\{1,\dots,n\}$. In that case, the marginal effect of increasing $q_{\bar t}$ and $\bar t$ such that the agent $q_{\bar t}$ remains indifferent has the following marginal effect on a more patient agent with $r<r(q_{\bar t})$:

\begin{align}\label{Aeq-prop3-start}
e^{-r \bar t}g(q_{\bar t})-re^{-r \bar t}G(q_{\bar t})\frac{\partial \bar t}{\partial q_{\bar t}}=e^{-r \bar t}g(q_{\bar t})\left(1-\frac{r}{r(q_{\bar t})}\right),
\end{align}
where we have used $\frac{\partial \bar t}{\partial q_{\bar t}}=\frac{g(q_{\bar t})}{r(q_{\bar t})G((q_{\bar t}))}$ from the indifference condition \eqref{eq-ICamount}. As \eqref{Aeq-prop3-start} is strictly positive, increasing the amount of experimentation makes all waiting agents strictly better off while keeping the experimenting agents equally well off. Hence, it is always optimal to increase $q_{\bar t}$ up to all agents with the same type experiment.

    To complete the proof, we provide examples of both cases: one where the amount of investment is larger under the designer's policy than under the social optimum and another where the opposite holds. For simplicity, let $n=2$. Suppose that $F_1$ is such that $x(F_{1})< x^\text{myop}$. Then, \Cref{prop:OptimalPureGN} and the above argument implies that the designer's policy must be such that good evidence is revealed after all high type agents have invested but before low types start to invest.
    
    We find examples where the first best calls for lower and higher experimentation than $F_1$ by considering the special case with $G(\theta)=1-e^{- \lambda_G \theta}$.
Then, it suffices to find examples where \eqref{eq:pureGN_FB_FOC} is strictly negative and positive at $F_1$, respectively.
To see that \eqref{eq:pureGN_FB_FOC} can be negative at $F_1$, take the limit as $\lambda_G\to\infty$ (the first term goes to $v_B<0$ and the second term goes 0). Hence, the designer's solution experiments more than the social optimum when the learning technology is good.

To see that the socially optimal investments can be strictly above $F_1$, notice that the second term in  \eqref{eq:pureGN_FB_FOC} goes to infinity as $f_2\to\infty$. Hence, the social optimum experiments more than the designer's solution when there are many patient agents.
\qed

\subsection{Proof of \Cref{prop:main2NEW}}\label{Assec-necessary}
The proof consists of a series of lemmas. 

The first one states that positive evidence that has not been disclosed at the beginning of the phase corresponding to agents of type $i$, will not be disclosed before all agents of type $i$ have invested. In particular, positive evidence generated by the investments of types $i$ will only be disclosed to later types. 
\begin{lemma}\label{lm:bunchingNEW}
      Under any optimal policy, ${H}^G$ is constant on  $(T_{i-1}, T_i)$ for $i\le \min\{\bar{n},n-1\} $.
\end{lemma}
\begin{proof}
Take a pair $(H, q)$ that is feasible and   
consider first types $i<\bar{n}$. 
$ T_i = \inf\{ t\ge 0 \colon  q_t > F_{i} \}$ is the time at which agents of type $i+1$ start experimenting under this policy. 
Suppose, contrary to the property in the lemma, that $H^G_{s'} > H^G_s$ for some $T_{i-1} < s < s' < T_{i} $.  
 We construct a new pair $(\hat{ H}, \hat{q})$ that increases welfare and has  $\hat{H}^G$  constant on $[ T_{i-1}, \hat{T}_i)$, where $\hat{T}_i$ is the time at which the new policy discloses all evidence that was originally disclosed in $( T_{i-1},  T_i]$ under the original policy. Thus, time $\hat{T}_i$ lets agents with type $r_i$ get the same expected payoff from waiting until time $ T_i$ under the original policy and from waiting until time $\hat{T}_i$ under the modified policy, and, thus, all $r_i$-agents experiment by $\hat{T}_i$ under the modified policy. 
Given a value $\hat{T}_i$, defined formally below, construct the modified $\hat{H}$ as  
 \begin{align*}
     \hat{H}^\w_t = \begin{cases}
         { H}_t^\w &\text{ for } t \le  T_{i-1}, 
         \\
         {H}_{ T_{i-1}}^\w &\text{ for } t \in ( T_{i-1}, \hat{T}_i),
          \\
         {H}_{t + ( T_{i}-\hat{T}_i)}^\w &\text{ for } t \ge \hat{T}_i, 
     \end{cases}& 
     & \text{ for } \w \in \{G,B\}.
 \end{align*}
The steps in the proof below will ensure that all agents experiment weakly earlier under the modified policy, guaranteeing feasibility.

The new disclosure time $\hat{T}_i$ is the time of an expected-utility-preserving contraction of all possible disclosures during $( T_{i-1},  T_i]$. 
For the probability of disclosure during this interval conditional on no disclosure before $T_{i-1}$ under the original policy define  $\hat h^G_i := \frac{ H^G_{ T_i} -  H^G_{ T_{i-1}}}{1-H^G_{ T_{i-1}}} >0 $ and $\hat h^B_i := \frac{ H^B_{ T_i} -  H^B_{ T_{i-1}}}{1-H^B_{ T_{i-1}}} \ge0 $. These will be the (conditional) probabilities of disclosure at $\hat T_i$ under the new policy.
Let $W_i$ denote the continuation utility from the original policy of type $i$ at time $ T_i$  if no evidence has been disclosed by time $ T_i$. By construction, the new policy gives the same continuation problem at time $\hat{T}_i$ as the original policy at time $ T_i$. Thus, $W_i$ will also be the time-$\hat{T}_i$ continuation utility under the new policy. 
Formally, $\hat{T}_i$ is defined by 
\begin{align*}
&e^{-r_i ( \hat{T}_i -  T_{i-1})}\Brac{x_{ T_{i-1}}v_G\hat h^G_i  + \Paren{ x_{  T_{i-1}} (1-\hat h^G_i) + (1-x_{ T_{i-1}})  (1-\hat h^B_i) } W_i} 
\\
= &x_{ T_{i-1}} v_G\int_{ T_{i-1}}^{ T_i} \frac{e^{-r_i (t-  T_{i-1})}  \de H^G_{t} }{1-H^G_{ T_{i-1}}}+   e^{-r_i ( T_i -  T_{i-1} ) } \Paren{ x_{  T_{i-1}}  (1-\hat h^G_i) + (1-x_{ T_{i-1}})  (1-\hat h^B_i) } W_i .
\end{align*}
The left-hand side is the expected utility under the new policy and the right-hand side under the original policy. 
Under the original policy, agents of type $i$ were (weakly) willing to invest at time $T_{i-1}$ rather than waiting until $T_i$. Thus, given the choice of $ \hat{T}_i$ above, those agents are still willing to invest at time $T_{i-1}$.
 This also implies that more impatient types $j<i$ are not waiting longer under the new policy than under the original policy, ensuring that the new policy is feasible.   

We now verify that the new policy strictly increases welfare. 
Divide both sides of the indifference condition above by the term in squared brackets to get 
\begin{align*}
e^{-r_i \hat{T}_i}
= &\frac{ x_{ T_{i-1}} v_G \int_{ T_{i-1}}^{ T_i} e^{-r_i t}  \frac{\de H^G_{t} }{1-H^G_{ T_{i-1}}} }{x_{ T_{i-1}}v_G\hat h^G_i  + \Paren{ x_{  T_{i-1}} (1-\hat h^G_i) + (1-x_{ T_{i-1}})  (1-\hat h^B_i) } W_i} 
\\
&+ e^{-r_i  T_i  } \frac{  \Paren{ x_{  T_{i-1}}  (1-\hat h^G_i) + (1-x_{ T_{i-1}})  (1-\hat h^B_i) } W_i}{x_{ T_{i-1}}v_G\hat h^G_i  + \Paren{ x_{  T_{i-1}} (1-\hat h^G_i) + (1-x_{ T_{i-1}})  (1-\hat h^B_i) } W_i}  .
\end{align*}
Below we make use of the fact that the term on the right-hand side is decreasing in the continuation utility  $ W_i$ because 
$
\int_{ T_{i-1}}^{ T_i} e^{-r_i t}  \frac{\de H^G_{t} }{1-H^G_{ T_{i-1}}}  >   e^{-r_i T_i  } \hat h^G_i $.

For all later types $k >i$, we show that the ex-ante expected utility from waiting until time $\hat{T}_i$ and then behaving optimally under the new policy strictly exceeds the payoff this type got from waiting until time $T_i$ and then behaving optimally  under the original policy. 
This is the case if and only if 
\begin{align} \label{eq: improvement cdt for hat T1_New}
 e^{-r_k\hat{T}_i }
  & > \frac{x_{T_{i-1}} v_G \int_{T_{i-1}}^{T_i} e^{-r_k t}  \frac{\de H^G_{t} }{1-H^G_{ T_{i-1}}}   }
 {x_{ T_{i-1}}v_G\hat h^G_i  + \Paren{ x_{  T_{i-1}} (1-\hat h^G_i) + (1-x_{ T_{i-1}})  (1-\hat h^B_i) } W_k} 
 \\
 &+ e^{-r_k T_i  }  \frac{  \Paren{ x_{T_{i-1}} (1-\hat h^G_i) + (1-x_{T_{i-1}}) (1-\hat h^B_i) }W_k}{x_{ T_{i-1}}v_G\hat h^G_i  + \Paren{ x_{  T_{i-1}} (1-\hat h^G_i) + (1-x_{ T_{i-1}})  (1-\hat h^B_i) } W_k}  .\notag
\end{align}
For the first term on the right-hand side we have 
$   \frac{\int_{T_{i-1}}^{T_i} e^{-r_k t}  \frac{\de H^G_{t} }{1-H^G_{ T_{i-1}}}  }{\hat h^G_i}   =   \frac{\int_{T_{i-1}}^{T_i} e^{-r_k t}  \de H^G_{t}}{ H^G_{T_i}-H^G_{T_{i-1}} } = \E \Brac{ e^{- r_k \tilde t} \ \lvert \ \tilde t \in (T_{i-1}, T_{i}]}$,
with the random arrival time of good news $\tilde t$. 
Use $r_k < r_i$ and the concavity of the function $(\cdot)^{r_k/r_i}$ to apply Jensen's inequality to this term. 
This shows that the right-hand side in \eqref{eq: improvement cdt for hat T1_New} is strictly smaller than
\begin{align*}
 & \Paren{\frac{x_{T_{i-1}} v_G \int_{T_{i-1}}^{T_i} \frac{e^{-r_i t}  \de H^G_{t} }{1-H^G_{ T_{i-1}}} }{x_{T_{i-1}}v_G\hat h^G_i} }^{r_k/r_i} 
 \frac{ x_{T_{i-1}} v_G \hat h^G_i}{ x_{T_{i-1}}v_G \hat h^G_i + \Paren{ x_{T_{i-1}} (1-\hat h^G_i) + (1-x_{T_{i-1}}) (1-\hat h^B_i) }W_k
 }  
 \\
 &+   \Paren{e^{-r_i  T_1 }}^{r_k/r_i}  \frac{  \Paren{ x_{T_{i-1}} (1-\hat h^G_i) + (1-x_{T_{i-1}}) (1-\hat h^B_i) }W_k }{x_{T_{i-1}}v_G \hat h^G_i + \Paren{ x_{T_{i-1}} (1-\hat h^G_i) + (1-x_{T_{i-1}}) (1-\hat h^B_i) }W_k}
\end{align*}
This is a weighted average of two terms, each elevated by an exponent smaller than 1. Apply Jensen's inequality again to conclude that the term is smaller than
\begin{align*}
 &\bigg(\frac{x_{T_{i-1}} v_G \int_{T_{i-1}}^{T_i} \frac{ e^{-r_k t}  \de H^G_{t} }{1-H^G_{ T_{i-1}}} }{x_{T_{i-1}}v_G\hat h^G_i} 
 \frac{ x_{T_{i-1}} v_G \hat h^G_i}{ x_{T_{i-1}}v_G \hat h^G_i + \Paren{ x_{T_{i-1}} (1-\hat h^G_i) + (1-x_{T_{i-1}}) (1-\hat h^B_i) }W_k
 }  
 \\
 &+   e^{-r_i  T_1 }  \frac{  \Paren{ x_{T_{i-1}} (1-\hat h^G_i) + (1-x_{T_{i-1}}) (1-\hat h^B_i) }W_k }{x_{T_{i-1}}v_G \hat h^G_i + \Paren{ x_{T_{i-1}} (1-\hat h^G_i) + (1-x_{T_{i-1}}) (1-\hat h^B_i) }W_k}\bigg)^{r_k/r_i}
\end{align*}
Note that $ W_k \ge W_i$ because $e^{-r_k t} > e^{-r_i t}$, so the more patient $r_k$-agents  can ensure at least the same continuation payoff as $r_i$-agents. We argued above that the term in parentheses above is decreasing in the continuation utility. Thus, the expression is bounded from above by the equivalent term where $W_k$ is replaced by $W_i$, which, by construction, equals $\Paren{e^{-r_i\hat{T}_i}}^{{r_k}/{r_i}} = e^{-r_k\hat{T}_i}$.
Hence, following the same strategy as under the original policy, agents with types $r_k < r_i$ get a strictly higher utility (i) because their expected payoff from receiving good news disclosed in $(T_{i-1}, T_{i}]$ under the original policy is increased by pooling the disclosures on time $\hat{T}_i$, and (ii) because $\hat{T}_i < T_i$ so that the no-disclosure continuation payoff $W_k $ accrues earlier.

Finally, consider type $i = \bar n$ when $\bar n <n$. That is, $i < n$ is the last type to experiment. 
Recall that $T_{\bar n }$ is the time at which the last agent experiments. We will show a stronger result than stated in the lemma: there is at most one time $\bar t \in  (T_{\bar n -1}, \infty)$ at which positive evidence can be disclosed.\footnote{Stronger because we may have $\bar t > T_{\bar n}$ so that $H^G$ is flat on a strictly larger interval than stated in the lemma.}  Formally, we show that there exists $\bar t$ such that \ 
$
    H^G_{\bar t -} = H^G_{T_{\bar n -1}+} \ \text{ and } \ H^G_{\bar t +} = H^G_{\infty}
$.

We follow similar steps as in the previous case with the difference that now all information that the original policy discloses on $(T_{\bar n -1}, \infty)$ is pooled on time $\bar t$. Time $\bar t$ is chosen such that the agents of type $\bar n$ are indifferent between experimenting at $T_{\bar n }$ under the original policy and waiting until $\bar t$ under the new policy. 
Since the modified policy discloses no information after $\bar t$, the continuation utility at time $\bar t$ is $v(\hat x_{\bar t})^+$ for any agent. 
Further $\bar n < n $ requires that under the original policy $\lim_{t\to \infty }v(x_t) \le 0$. Since all information that was ever disclosed under the original policy will be disclosed at $\bar t$ under the new policy we conclude that $v(\hat x_{\bar t})^+ = \Paren{\lim_{t\to \infty }v(x_t)}^+ = 0$. 
 Thus, we choose $\bar t$ to satisfy 
\begin{dense}
    \begin{align}\label{eq:proof-L1-bart_New}
e^{-r_{\bar n}  \bar t } x_{ T_{\bar n-1}}v_G \bar{h}^G_{\bar n}   &=
x_{ T_{\bar n-1}} v_G\int_{ T_{\bar n-1}}^{ T_{\bar n}} \frac{e^{-r_{\bar n} t}  \de H^G_t}{1- H^G_{T_{\bar n-1}}} 
\\
&
+   e^{-r_{\bar n}  T_{\bar n} } \Paren{ x_{  T_{\bar n-1}}  (1- \hat h^G_{\bar n})+ (1-x_{ T_{\bar n-1}})  (1-\hat h^B_{\bar n}) } W_{\bar n} , \notag
\end{align}
\end{dense}
\noindent where $\bar{h}^\w_{\bar n} := \frac{H^\w_\infty - H^\w_{T_{\bar n-1}}}{1-H^\w_{T_{\bar n-1 }}} $, and $\hat h^\w_{\bar n}= \frac{H^\w_{T_{\bar n}} - H^\w_{T_{\bar n-1}}}{1-H^\w_{T_{\bar n-1}}}$. 
As in the previous case, the maintained willingness of agents of type $\bar n$ to invest at $T_{\bar n-1}$ implies that more impatient agents are not induced to wait longer by the modification, ensuring feasibility.

We show that, with $\bar t$ in \eqref{eq:proof-L1-bart_New}, the more patient types $k > \bar n$ are strictly better off waiting forever under the new policy than waiting forever under the original policy, i.e.: 

\vspace{-1cm}

\begin{align}\label{eq:lm1-case2-improvement}
    e^{-r_k \bar t}  > \frac{ \int_{T_{\bar n-1}}^\infty e^{- r_k t} \frac{ \de H^G_t}{1-H^G_{T_{\bar n -1}}} }{\bar{h}^\w_{\bar n} }, \quad \text{ for all } k > \bar n.
\end{align}
By \eqref{eq:proof-L1-bart_New}, $ e^{- r_k \bar t}$ is equal to
 \begin{dense}
    \begin{align*} 
& \Big ( \frac{x_{ T_{\bar n-1}} v_G\int_{ T_{\bar n-1}}^{ T_{\bar n}} \frac{e^{-r_{\bar n} t}  \de H^G_t}{1- H^G_{T_{\bar n-1}}} }{x_{ T_{\bar n-1}}v_G \bar{h}^G_{\bar n}}
+  \frac{e^{-r_{\bar n}  T_{\bar n} } \Paren{ x_{  T_{\bar n-1}}  (1- \hat h^G_{\bar n})+ (1-x_{ T_{\bar n-1}})  (1-\hat h^B_{\bar n}) } W_{\bar n}}{x_{ T_{\bar n-1}}v_G \bar{h}^G_{\bar n}} \Big )^{r_k / r_{\bar n}}
\end{align*}
\end{dense}

Once having reached  $T_{\bar n}$, agents of type $\bar n$ always have the option to wait forever, thus their continuation utility satisfies 
$
    W_{\bar n} \ge  x_{T_{\bar n }} v_G \int_{T_{\bar n }}^\infty \frac{e^{-r_{\bar n }(t-T_{\bar n })} \de H^G_t}{1- H^G_{T_{\bar n-1}}} . 
$
Hence, 
 $ e^{- r_k \bar t}$ is at least 
 \begin{dense}
    \begin{align*} 
& \bigg( \frac{x_{ T_{\bar n-1}} v_G\int_{ T_{\bar n-1}}^{ T_{\bar n}} \frac{e^{-r_{\bar n} t}  \de H^G_t}{1- H^G_{T_{\bar n-1}}} }{x_{ T_{\bar n-1}}v_G \bar{h}^G_{\bar n}}
\\
&+  \frac{e^{-r_{\bar n}  T_{\bar n} } \Paren{ x_{  T_{\bar n-1}}  (1- \hat h^G_{\bar n})+ (1-x_{ T_{\bar n-1}})  (1-\hat h^B_{\bar n}) } x_{T_{\bar n }} v_G \int_{T_{\bar n }}^\infty \frac{e^{-r_{\bar n }(t-T_{\bar n })} \de H^G_t}{1- H^G_{T_{\bar n-1}}}}{x_{ T_{\bar n-1}}v_G \bar{h}^G_{\bar n}} \bigg)^{r_k / r_{\bar n}}
\end{align*}
\end{dense}
Note that $\Paren{ x_{  T_{\bar n-1}}  (1- \hat h^G_{\bar n})+ (1-x_{ T_{\bar n-1}})  (1-\hat h^B_{\bar n}) } x_{T_{\bar n }} = 
x_{T_{\bar n -1}}  (1- \hat h^G_{\bar n})$. 
Thus, $ e^{- r_k \bar t}$ is at least 
$ \Big ( \frac{x_{ T_{\bar n-1}} v_G\int_{ T_{\bar n-1}}^{\infty } e^{-r_{\bar n} t}   \frac{\de H^G_t}{1- H^G_{T_{\bar n-1}}}}{x_{ T_{\bar n-1}}v_G \bar{h}^G_{\bar n} } \Big )^{r_k / r_{\bar n}}.
$ 
Since $r_k/r_{\bar n } < 1$, Jensen's inequality implies that the above is greater than the right-hand side of \eqref{eq:lm1-case2-improvement}. 
This proves that more patient agents are strictly better off. 
\end{proof}


Building on \Cref{lm:bunchingNEW}, we show that any optimal policy must disclose the generated bad evidence before the next type of agents start to invest:
\begin{lemma}\label{prop:BNNEW}
    Under any optimal policy, ${H}_{T_i}^B\geq B(F_i)$ for all $i<\bar{n}$.
\end{lemma}

\vspace{-0.4cm}

\noindent \textit{Proof.}  Suppose that $q$ is incentive compatible given $(H^B,H^G)$,  ${H}_t^G$ is constant for all $t\in (T_{i-1}, T_i)$ if $i<\bar{n} $ as in \Cref{lm:bunchingNEW}, and, contrary to the claim,   
$H_{T_i}^B<B(F_i)$ for some $i<\bar{n} $. 
We show that there exists another pair $(\hat{H}, \hat{q})$ that satisfies the IC constraint, has ${H}_{T_i}^B\geq B(F_i)$ for all $i<\bar{n}$, and leads to strictly higher  welfare.

We apply an induction argument. First, we show that if the original policy does not disclose all bad evidence at $T_{\bar{n} -1}$, there is a strict improvement that does. Then, we show that for all $i < \bar n$, there is a strict improvement if not all bad evidence is disclosed at $T_{i-1}$ but all bad evidence is disclosed at $T_{i}$.

Let $\lim_{t\to\infty}H_t^{\omega}=: H^{\omega}_{\infty}$.
As a preliminary step, if  $H^B_{\infty}<B(F_{\bar{n} })$, $\bar{n}<n$, and $\frac{x_{0}(1-H_{\infty}^G)}{x_{0}(1-H_{\infty}^G)+(1-x_0)(1-B(F_{\bar{n}}))}\geq x^\text{myop}$, then it would be a strict improvement to disclose the additional available bad news at some time $t$ chosen such that types $\bar{n} +1$ are willing to experiment at $t$ without inducing more impatient agents to wait until $t$. Hence, we focus on the cases where either $\bar{n} =n$ or $\frac{x_{0}(1-H_{\infty}^G)}{x_{0}(1-H_{\infty}^G)+(1-x_0)(1-B(F_{\bar{n}}))}<x^\text{myop}$.

Suppose that  ${H}_{T_{\bar{n} -1}}^B<B(F_{\bar{n} -1})$. 
 We suggest an alternative policy $\hat{ H}$ that is otherwise the same as the original but has ${\hat{ H}}^B_{t}=B(F_{\bar{n} -1})$ for all $t\in [\hat T_{\bar n -1}, T_{\bar{n} })$, where time $\hat T_{\bar n -1} >T_{\bar{n} -1}$ is chosen such that type $\bar{n} -1$ is indifferent between investing at $T_{\bar{n} -1}$ and waiting until the additional bad evidence is disclosed at $\hat T_{\bar n -1}$. The fact that type $r_{\bar{n} -1}$ is indifferent implies that more patient agents $\bar{n} $ are strictly better off waiting until $\hat T_{\bar n -1}$ and all less patient agents  $r>r_{\bar{n} -1}$  strictly prefer investing at time $T_{\bar{n} -1}$ or earlier.
As the incentives of the less patient agents remain unchanged, it remains incentive compatible to have $\hat H_t^{\omega} = H_t^{\omega}$ for all $t\leq T_{\bar{n} -1}$, and hence, the expected payoff of types $ i <\bar{n} $ remains unchanged. Type $\bar{n} $ is strictly better off.

To complete the first step, we need to check that the modified policy implements the same good news process as the original policy. If $\bar n =n$, any good news released after $T_{\bar n-1}$ is irrelevant because all remaining agents invest also absent news. Hence, we focus on $\bar{n} <n$.  Let $\bar{t}:=\inf\{t>T_{\bar{n} -1}: H_t^G>H^G_{T_{\bar{n}-1}}\}$ be the earliest time after $T_{\bar{n} -1}$ at which the original policy discloses good news. 
The amount of good news generated by $\bar t$ under the original policy will also be available under the modified policy if $\hat T_{\bar n -1} \le \bar{t}$ because, by construction, agents of type $\bar{n} $  prefer investing at $\hat T_{\bar n -1}$ to waiting beyond $\bar{t}$. 
To see this, recall that types $\bar{n} $ weakly prefer investing at $T_{\bar{n} -1}$ to waiting until $\bar t$ under the original scheme, and investing at $\hat T_{\bar n -1}$ under the alternative scheme is strictly better for them.

Hence, we are left to verify that $\hat T_{\bar n -1} \le \bar{t}$,  
This is immediate if $ \bar{t}=\infty$.   
For the case of $\bar{t}<\infty$, we know from the proof of  \Cref{lm:bunchingNEW} that a one-time disclosure of good news is strictly better than multiple disclosures after $T_{\bar{n}-1 }$, which implies that we can focus on the case where $H^{G}_{\infty}=H^{G}_{\bar{t}}$.\footnote{There always exists a policy that is otherwise identical to the original policy but pools all good-news disclosure times after $T_{\bar{n}-1 }$ and gives an upper bound for the welfare under the original policy.}  
This tells us that $x^\text{myop}>\frac{x_{0}(1-H_{\infty}^G)}{x_{0}(1-H_{\infty}^G)+(1-x_0)(1-B(F_{\bar{n}}))}\geq\frac{x_{0}(1-H_{\bar t}^G)}{x_{0}(1-H_{\bar t}^G)+(1-x_0)(1-B(F_{\bar{n}-1}))}$, which implies that $x_{0}(1-H_{\bar t}^G)v_G+(1-x_0)(1-B(F_{\bar{n}-1}))v_B<0$.

As investing at $T_{\bar{n} -1}$ is incentive compatible under the original scheme, we have
\begin{align}\label{Aeg-lm2x}
    v(x_{T_{\bar{n} -1}})\geq e^{-r_{\bar{n} }(\bar{t}-T_{\bar{n} -1})}x_{T_{\bar{n} -1}}\hat{H}^G_{\bar{n}}v_G,
\end{align}
where $\hat H^G_{\bar{n} }:=H^G_{\bar{t}}-H^G_{T_{\bar{n} -1}}$.
By the construction of $\hat T_{\bar n -1}$, we also have
\begin{align}\label{Aeg-lm2y}
    v(x_{T_{\bar{n} -1}})< e^{-r_{\bar{n} }(\hat T_{\bar n -1}-T_{\bar{n} -1})}\left(x_{T_{\bar{n} -1}}v_G+(1-x_{T_{\bar{n} -1}})(1-\hat H_{\bar{n}}^B)v_B\right),
\end{align}
where $\hat H^B_{\bar{n} }:=B(F_{\bar{n} -1})-H^B_{T_{\bar{n} -1}}$.
Combining \eqref{Aeg-lm2x} and \eqref{Aeg-lm2y} and rearranging gives
\begin{align*}
    e^{-r_{\bar{n} }(\bar{t}-\hat T_{\bar n -1})}<\frac{x_{T_{\bar{n} -1}}(1-\hat H^G_{\bar{n}})v_G+(1-x_{T_{\bar{n} -1}})(1-\hat H^B_{\bar{n}})v_B}{x_{T_{\bar{n} -1}}\hat H^G_{\bar{n}}v_G}+1,
\end{align*}
where the first term on the RHS is negative by $x_{0}(1-H_{\bar t}^G)v_G+(1-x_0)(1-B(F_{\bar{n}-1}))v_B<0$. Hence, $\hat T_{\bar n -1}<\bar{t}$.

The second step of the induction argument is almost identical.  Suppose no full revelation of bad news at $T_{i-1}$ but full revelation of bad news at $T_{i}$. Define an alternative policy that discloses all bad evidence from $r\geq r_{i-1}$ agents at $\hat T_{i-1} \in(T_{i-1}, T_{i})$ such that type $i-1$ is indifferent between investing at $T_{i-1}$ and waiting until $\hat T_{i-1}$. All more patient types are better off as they strictly prefer waiting, and the incentives and expected payoff for all more impatient types remain unchanged. To see that $\hat T_{i-1}<T_i$, recall that we assume full bad news revelation at $T_i$. Hence, there is more information available at $T_i$ than at $\hat T_{i-1}$. Since the original scheme is incentive compatible, $i-1$ must prefer investing at $T_{i-1}$ over waiting until $T_i$, which then gives that waiting until $\hat T_{i-1}$ under the alternative scheme must be better than waiting until $T_i$, which is only possible if $\hat T_{i-1}<T_i$.
 \qed 


Together with the feasibility constraint that $H_t^B\leq B(q_{t-})$ \Cref{prop:BNNEW} implies that $H_t^B= B(F_i)$ for all $i<\bar{n}$.  The following two lemmas follow from \Cref{prop:BNNEW}: 
\begin{lemma}\label{lm:crossingNEW}
Under any optimal policy, the no-disclosure belief $x_t$ crosses $x^\text{myop}$ at most once.
\end{lemma}

\vspace{-0.4cm}

\noindent \textit{Proof.} Suppose not. Then it must be that $x_{t}<x^\text{myop}$ for some $t < T_{\bar{n} -1}$, where $\bar{n} $ is the last type who invests if no news is revealed. Since for $i<\bar{n}$ good news are only revealed at times $t=T_i$ for some $i$ (\Cref{lm:bunchingNEW}), this requires $x_{T_i}<x^\text{myop}$ for some $i< \bar{n}$. From \Cref{prop:BNNEW}, we further know that $H_{T_i}^B=B(F_i)$ for all $i<\bar{n}$. But then no further agent is willing to invest given the current belief,  and, since all generated bad news has been disclosed already,  it is impossible to have $x_t\geq x^\text{myop}$ for any $t>T_i$. The fact that $x_t< x^\text{myop}$ for all $t>T_i$ contradicts $i<\bar{n} $.
\qed

\begin{lemma}\label{lm:hat iNEW} Under any optimal policy, a)  $\bar{n} = n$ if $x(F_{n-i})> x^\text{myop}$ for all $i\in\{1,\dots, n\}$, and b) $\bar{n} < n$ if $x(F_{n-1})< x^\text{myop}$.
\end{lemma}

\vspace{-0.4cm}

\noindent \textit{Proof.} Part (a) is an immediate consequence of Lemmas \ref{lm:bunchingNEW} and \ref{prop:BNNEW} as the public belief always stays above the myopic threshold.

 We prove part (b) by arguing that it is strictly better to reveal all good news at $T_{n-1}$ rather than hiding enough good news to make type $n$ willing to invest absent news.  Recall that all agents with type $n$ receive the same expected payoff as the first of them to invest.  Hence, fixing disclosure and investment choices prior to $T_{n-1}$, the question of optimality boils down to maximizing type $n$'s welfare at time $T_{n-1}$. 
Suppose, contrary to the statement, that there is an optimal policy such that $x_{T_{n-i}}\geq x^\text{myop}$ for all $i\in\{1,\dots, n\}$. We suggest that there is a strict improvement to reveal all good news with a long enough delay, $\Delta$, chosen so that type $n-1$ is indifferent between investing at $T_{n-1}$ and waiting until the revelation of good news at $T_n + \Delta$.\footnote{\linespread{1.0}\selectfont Such a $\Delta$ exists when $x_{T_{n-2}}>x^\text{myop}$. To rule out $x_{T_{n-2}} = x^\text{myop}$, notice that it would imply that no news was ever revealed because otherwise all agents would be better off waiting for the news rather than investing with a payoff of $v(x^\text{myop}) = 0$, contradicting incentive compatibility. In that case, one can use the same argument as in this proof, but applied to type $n-2$ instead of $n-1$. If also $x_{T_{n-3}} = x^\text{myop}$, continue moving further until reaching type $1$ for which we have $x_{T_{0}}=x_0> x^\text{myop}$.} 
If $n-1$ is indifferent, agents with type $r_n < r_{n-1}$ must strictly prefer to wait for the good news at $T_n + \Delta$ and are hence strictly better off than under the original policy. Thus no policy featuring $\bar{n} = n$ is optimal when $x(F_{n-1}) < x^\text{myop}$. 
\qed


To complete the proof of \Cref{prop:main2NEW}, we show the necessity of good-news delays \textit{across} types.
All agents with the same discount rate must get the same expected payoff as the agent with that discount rate who experiments first. Therefore, we can rewrite the designer's problem \eqref{eq:MaxProblem_general} using the investment time $T_{i-1}$ to compute the expected payoff of all type-$i$ agents and specify the amount of disclosed evidence only at times in  $\{T_i\}_{i\ge 0}$. The incentive compatibility constraints simplify to requiring that agents of type $i < \bar n$ weakly prefer investing at time $T_{i-1}$ to waiting until time $T_i$, and agents of type $\bar n $ prefer investing at $T_{\bar n-1}$ to waiting until $\bar t$, where $\bar{t}:=\inf\{t>T_{\bar{n} -1}: H_t^G>H^G_{T_{\bar{n}-1}}\}$ is the earliest time at which good news is disclosed after  $T_{\bar{n} -1}$. By \Cref{lm:bunchingNEW}, there is no good-news disclosure within a phase. \Cref{prop:BNNEW} implies that $H_{T_i}^B=B(F_i)$. Hence, the value of the designer's problem can be determined by the following finite-dimensional  problem, where the maximization is over the last experimenting type  $\bar n $, the times $\{T_i\}_{i=1}^{\bar n-1}$ and $\bar{t}$,  and the amount of good news disclosed at those times $\{H^G_i\}_{i=1}^{\bar{n}}$:

\vspace{-1cm}

\begin{dense}  
\begin{multline}\label{Aeq-lemma5}
\begin{aligned}
   \max    \; \Big\{
   &\sum_{i=1}^{\bar{n}-1}x_0 v_G \Paren{(H^G_i-H^G_{i-1})\sum_{j=i+1}^n f_j e^{-r_j T_i} +  \Paren{H^G_{\bar n} - H^G_{\bar n -1}}\sum_{j=\bar n+1}^n f_j e^{-r_j \bar t }}
  \\&    + \sum_{i=1}^{\bar{n}} f_i e^{-r_i T_{i-1}}\Paren{x_0 v_G (1-H^G_{i-1}) +(1-x_0)v_B (1- B(F_{i-1}))} \Big\}
\\
 \text{ such}&\text{ that  $ \forall i\le \bar n \colon  H_{i-1}^G\le H_i^G \le G(F_i)$, and $\forall i  < \bar{n}$,}
     \\
    &    e^{-r_i T_{i-1}} \Paren{x_0 v_G (1-H^G_{i-1}) + (1-x_0)v_B (1-B(F_{i-1}))} 
    \\
    & \ge   e^{-r_i T_i}{\Big [}x_0 v_G (H^G_i - H^G_{i-1}) +  \Paren{x_0 v_G (1-H^G_{i}) + (1-x_0)v_B (1-B(F_i))  }^+ {\Big ]} 
     \\ 
  \text{ and }
  &    e^{-r_{\bar n } T_{\bar n-1}} \Paren{x_0 v_G (1-H^G_{\bar n -1}) + (1-x_0)v_B  (1-B(F_{\bar n -1}))} 
    \\
    & \ge   e^{-r_{\bar n } \bar t }{\Big [}x_0 v_G (H^G_{\bar n } - H^G_{\bar n -1}) +  \Paren{x_0 v_G (1-H^G_{\bar n}) + (1-x_0)v_B (1-B(F_{\bar n }))  }^+ {\Big ]}.  
\raisetag{5em}
   \end{aligned}
\end{multline}
\end{dense}

Recall the convention that  $T_0 = H^G_0=0$. Define for ease of exposition $\tilde V_{i} := x_0 v_G (1-H^G_{i-1}) +(1-x_0)v_B (1-B(F_{i-1})) $ and $\hat V_{i} :=  x_0 v_G (1-H^G_{i-1}) +(1-x_0)v_B (1-B(F_{i})) $ for the terms on the left- and on the right-hand side of the IC constraints, respectively. Note that they differ from $v$ because they are ex-ante expected values whereas $v(x_{T_{i-1}})$ is \textit{conditional} on reaching time $T_{i-1}$ without evidence. The value  $\tilde V_i$ considers all evidence disclosed at $T_{i-1}$ whereas $\hat V_i$ also takes into account the (lack of) bad evidence to be disclosed at $T_i$. 

By  \Cref{lm:crossingNEW} and the definition of $\bar{n}$, $x_t > x^\text{myop}$ for all $t\le T_{\bar{n}-1}$, so that the last parenthesis on the right-hand side is positive. 
Thus, for $i<\bar{n}$ the right-hand side of the IC constraint equals $ e^{-r_i T_i}[x_0 v_G (1-H^G_{i-1}) +(1-x_0)v_B (1-B(F_{i}))]=  e^{-r_i T_i} \hat V_{i}$. For $i=\bar{n}$, the right-hand side of the constraint equals $e^{-r_{\bar{n}} \bar t }x_0 v_G (H^G_{
\bar n} - H^G_{\bar n -1}) > 0$ when $\bar{n} <n $. We focus on the case $\bar{n} <n $. The proof for $\bar{n} = n$ is parallel.
Assigning Lagrange-multiplier $\gamma_i$ to the constraint with $r_i$, the first-order condition (FOC) with respect to $\bar t$ is 
\begin{align}\label{eq:FOC_T_hat_i}
  x_0 v_G (H^G_{\bar n}-H^G_{\bar n -1}) r_{\bar{n} }\Brac{-  \sum_{j=\bar{n}+1}^n f_j \frac{r_j}{r_{\bar{n} }}  e^{-r_j \bar t } + \g_{\bar{n} }  e^{-r_{\bar{n}} \bar t }} = 0.
\end{align}
Since $r_j < r_{\bar{n} }  $ for all $j>\bar{n}$ and $H^G_{\bar n} > H^G_{\bar n -1}$, we have $ \sum_{j=\bar{n}+1}^n f_j e^{-r_j \bar t } > \g_{\bar{n}}  e^{-r_{\bar{n}} \bar t }  $.
The derivative with respect to $H^G_{\bar{n}}$ is 
$
 x_0 v_G \Brac{  \sum_{j=\bar{n}+1}^n f_j e^{-r_j \bar t } - \g_{\bar{n}} e^{-r_{\bar{n}}\bar t } }.
$
Therefore, $H^G_{\bar{n}}$ must be equal to the upper bound $G(F_{\bar{n}})$.

Next, consider the derivative with respect to $H^G_{\bar n -1}$. This is $x_0 v_G >0$ multiplied with 
\begin{align}
 \sum_{j=\bar n }^n f_j e^{-r_j T_{\bar n-1}}    - \sum_{j=\bar n +1}^n f_j e^{-r_j \bar t}  - f_{\bar n } e^{-r_{\bar n } T_{\bar n-1}} - \g_{\bar n } ( e^{-r_{\bar n } T_{\bar n -1}} -e^{-r_{\bar n } \bar t} ).
\end{align}
We show first that this derivative is negative and then show by induction that the derivative with respect to $H^G_i$ is negative for all $i<\bar{n}-1$.

For the first step  $i=\bar{n} -1$ we have
$
   \g_{\bar{n} } =  \sum_{j=\bar{n}+1}^n f_j \frac{r_j}{r_{\bar{n} }} \frac{ e^{-r_j \bar t }}{ e^{-r_{\bar{n}} \bar t }} >  \sum_{j=\bar{n}+1}^n f_j \frac{e^{-r_j T_{\bar{n}-1}}-e^{-r_j \bar t }}{e^{-r_{\bar{n}} T_{\bar{n}-1}} -e^{-r_{\bar{n}} \bar t } }, $
where the equality  follows from  \eqref{eq:FOC_T_hat_i}, and the inequality holds because the relation between the summation terms is equivalent to $
\frac{e^{r_{\bar{n}} (\bar t -T_{\bar{n}-1})} -1 }{r_{\bar{n} }} > \frac{e^{r_j(\bar t - T_{\bar{n}-1})}-1}{r_j}
$, and  $r_{\bar{n}}>r_j$ for $j>\bar n$. Thus, the derivative with respect to $H^G_{\bar n -1}$ is negative, so that $H^G_{\bar n -1} - H^G_{\bar n -1}$. 

For $i<\bar{n}-1$, the derivative with respect to $H^G_i$  is $x_0 v_G >0$ times
\begin{align}
 \sum_{j=i+2}^n f_j e^{-r_j T_{i}}     - \sum_{j=i+2}^n f_j e^{-r_j T_{i+1}}  - \g_{i+1} ( e^{-r_{i+1} T_{i}} -e^{-r_{i+1} T_{i+1}} ).
\end{align}
We show by induction that the derivative above is negative for all $i<\bar{n}-1$. That is, going from higher to lower $i$, we show that
\begin{align}\label{eq:gamma_i large}
  \g_{i+1} >    \sum_{j=i+2}^n f_j \frac{e^{-r_j T_{i}}-e^{-r_j T_{i+1}}}{e^{-r_{i+1} T_{i}} -e^{-r_{i+1} T_{i+1}} } \quad \text{implies} \quad  \g_{i} >    \sum_{j=i+1}^n f_j \frac{e^{-r_j T_{i-1}}-e^{-r_j T_{i}}}{e^{-r_{i} T_{i-1}} -e^{-r_{i} T_{i}} } .
\end{align}

The FOC with respect to $T_{ i}$ for $i<\bar{n}-1$ gives that the following must equal zero:

\vspace{-1cm}

\begin{dense}
    \begin{align*}
 & -x_0 v_G (H^G_{ i}-H^G_{ i-1})  \sum_{j= i+1}^n f_j {r_j} e^{-r_j T_{ i}} - f_{i+1}r_{i+1} e^{-r_{i+1}T_i} \tilde V_{i+1} 
 + \g_{i} r_i e^{-r_{i} T_{i}}\hat V_{i} - \g_{i+1} r_{i+1}e^{-r_{i+1} T_{i}}\tilde V_{i+1}  ,
\end{align*}
\end{dense}

\vspace{-0.6cm}

\noindent which is equivalent to
$
  \g_i = $ $ x_0 v_G (H^G_{ i}-H^G_{ i-1})  \sum_{j= i+1}^n f_j \frac{{r_j} e^{-r_j T_{ i}}}{ r_i e^{-r_{i} T_{i}}\hat V_{i}} +$  $ f_{i+1} \frac{r_{i+1} e^{-r_{i+1}T_i} \tilde V_{i+1}}{ r_i e^{-r_{i} T_{i}}\hat V_{i}}$ \\ $ 
 + \g_{i+1} \frac{r_{i+1}e^{-r_{i+1} T_{i}}\tilde V_{i+1}}{ r_i e^{-r_{i} T_{i}}\hat V_{i}}.$
We have 
$
\frac{\tilde V_{i+1}}{\hat V_{i}} = \frac{x_0 v_G (1-H^G_{i}) +(1-x_0)v_B(1-B(F_i)) }{x_0 v_G (1-H^G_{i-1}) +(1-x_0)v_B (1-B(F_i))}
$.
Note that for $i <\bar{n}$, the derivative with respect to $H^G_{i}$ is negative when $\g_{i+1}$ satisfies the left part \eqref{eq:gamma_i large}, so that   $H^G_{i}=H^G_{i-1}$. Then,  $\tilde V_{i+1}/\hat V_i =1 $. 
Consequently, $\gamma_i $ satisfies the right part of \eqref{eq:gamma_i large} if  
\begin{dense}
\begin{align*}
   f_{i+1} \frac{r_{i+1} e^{-r_{i+1}T_i}}{ r_i e^{-r_{i} T_{i}}} +  \sum_{j=i+2}^n f_j \frac{e^{-r_j T_{i}}-e^{-r_j T_{i+1}}}{e^{-r_{i+1} T_{i}} -e^{-r_{i+1} T_{i+1}} }  \frac{r_{i+1}e^{-r_{i+1} T_{i}}}{ r_i e^{-r_{i} T_{i}}} 
   \ge  \sum_{j=i+1}^n f_j \frac{e^{-r_j T_{i-1}}-e^{-r_j T_{i}}}{e^{-r_{i} T_{i-1}} -e^{-r_{i} T_{i}} }. 
\end{align*}
\end{dense}
For the $j=i+1$ term we have again $(e^{r_i(T_i-T_{i-1})}-1)/r_i > (e^{r_{i+1}(T_i-T_{i-1})}-1)/r_{i+1} $, and for the terms $j\ge i+2$ we have\footnote{This inequality is equivalent to $$\frac{e^{r_i(T_i-T_{i-1})}-1}{r_i}{\Big /}\frac{e^{r_j (T_i-T_{i-1})}-1}{r_j}
>
\frac{1-e^{-r_{i+1}(T_{i+1}-T_{i})}}{r_{i+1}}{\Big /} \frac{1-e^{-r_j (T_{i+1}-T_i)}}{r_j },$$
where the left-hand sides is greater than 1 and the right-hand side is smaller than 1 for $r_j <r_{i+1}<r_i$.} 
\begin{align*}
\frac{e^{-r_j T_{i}}-e^{-r_j T_{i+1}}}{e^{-r_{i+1} T_{i}} -e^{-r_{i+1} T_{i+1}} }  \frac{r_{i+1}e^{-r_{i+1} T_{i}}}{ r_i e^{-r_{i} T_{i}}} >\frac{e^{-r_j T_{i-1}}-e^{-r_j T_{i}}}{e^{-r_{i} T_{i-1}} -e^{-r_{i} T_{i}} }.
\end{align*}
It follows that $\g_i$ is sufficiently large, and any solution to the above problem has $H^G_i =0$ for all $i <\bar{n}$; and, whenever $\bar{n} <n$, then $H^G_{\bar{n}} =  G(F_{\bar{n}})$.
 \qed

 \subsection{Proof of \Cref{prop:main1NEW}}\label{Asec:main1NEW}

The proofs of Lemmas \ref{lm:bunchingNEW}-\ref{lm:crossingNEW} show that if a policy does not satisfy the property in the corresponding result statement, then there is another policy that satisfies the property and delivers strictly higher welfare. Therefore, we know that the designer's problem \eqref{eq:MaxProblem_general} is equivalent to the problem where we restrict to policies satisfying the properties of Lemmas \ref{lm:bunchingNEW}-\ref{lm:crossingNEW}. Importantly, this holds for the supremum in Problem \eqref{eq:MaxProblem_general}, even if the maximum was not attained.

Combining the properties of Lemmas \ref{lm:bunchingNEW}-\ref{lm:crossingNEW}, we can restrict attention to policies where, for all $i<\bar{n}\colon$ $H_t^G$ changes at most at times  $\{T_i\}$;  ${H}_{t}^B\geq B(F_i)$ for all $t>T_i$; and where the no-disclosure belief crosses $x^{myop}$ at most once. Furthermore, any incentive compatible policy must give all agents of the same type get the same expected payoff. This implies that problem \eqref{eq:MaxProblem_general} is equivalent to the problem \eqref{Aeq-lemma5} in the proof of \Cref{prop:main2NEW}. We can then replace sup with max in the designer's problem as \eqref{Aeq-lemma5} is a finite-dimensional maximization problem with continuous objective on a compact domain.\footnote{While $T_i$ could in principle take any values in $\mathbb{R}_+$, the relevant values of $\{T_i\}$ can easily be restricted to a sufficiently large compact set as the objective is eventually decreasing in $T_i$.}

We are left to show that for any policy that satisfies the necessary conditions, there exists another policy that takes the form described in \Cref{prop:main1NEW}, is incentive compatible, and achieves at least the same welfare.
The good-news part of \Cref{prop:main1NEW} is immediate: Any solution to \eqref{Aeq-lemma5} must have ${H}_{T_i}^G=0$ for all $i<\bar{n} $ and ${H}_{t}^G = G(F_{\bar{n}}) $ for all $t >  T_{\bar{n}}$ if $\bar{n} <n$.
Use  $T_{\bar{n}}$ as the $\bar t$ in the statement of \Cref{prop:main1NEW}. Recall $\lim_{t\to\infty}q_t=F_{\bar{n}} $. 
 Furthermore, a potential good-news revelations in $(T_{\bar{n} -1}, T_{\bar{n}})$, which does not make the no-disclosure belief drop below the myopic threshold, does not affect the objective or the IC constraints in \eqref{Aeq-lemma5}. Therefore, the designer can equally well set ${H}_{t}^G=0$ also for $t\in (T_{\bar{n} -1 }, T_{\bar{n}})$.

  To show the bad news part of \Cref{prop:main1NEW}, we argue that in any policy that maximizes \eqref{Aeq-lemma5}, the period length $T_i-T_{i-1}$ is such that type $i$ is indifferent between experimenting at $T_{i-1}$ and waiting until $T_i$ for all $i<\bar{n} $. 
Suppose instead that type $i$ strictly prefers experimenting at $T_{i-1}$ to waiting until $T_i$. Then, it is a strict improvement to give out bad news already at $T_i-\epsilon$: type $i+1$ is strictly better off and the IC for type $i$ is still satisfied for $\epsilon$ small enough. 

 With each type $i<\bar{n} $ indifferent between experimenting at $T_{i-1}$ and waiting until $T_i$ and because any optimal scheme reveals all generated bad news by $T_i$, we have:
  \begin{align*}
      x_{T_{i-1}}v_G+(1-x_{T_{i-1}})v_B=e^{-r_i(T_i-T_{i-1})}x_{T_{i-1}}v_G+(1-x_{T_{i-1}})\frac{B(F_t)-B(F_{t-1})}{1-B(F_{t-1})}v_B.
  \end{align*}
This is the same condition as under transparent bad news. Hence, the implied disclosures at any $T_i$ for $i<\bar{n} $ under transparent bad news are the same as and under any scheme solving \eqref{Aeq-lemma5}. Since all agents of the same type must get the same welfare, bad-news disclosures within $(T_{i-1}, T_i)$ do not affect welfare. 

Finally, consider the last phase, $(T_{\bar{n} -1},T_{\bar{n} })$. If $\bar{n} =n$, the disclosure policy after $T_{\bar{n} -1}$ does not affect anyone's payoff, and hence transparent bad news is trivially optimal. Let $\bar{n} <n$. 
Since bad news is not decision relevant for $i>\bar{n} $, transparency does not affect their payoffs. Furthermore, it does not affect the IC of type $\bar{n}$ because the rate of investment adjusts so that type $\bar{n}$ remains indifferent.

We conclude that for all optimal policies, there exists a policy that takes the form of the policy in the claim and gives the same expected payoff for each type as the original policy. Hence, the altered policy is also optimal.
 \qed 


  \subsection{Proof of \Cref{lm:Pareto}}\label{Asec_ParetoProof}
The proofs of Lemmas \ref{lm:bunchingNEW}-\ref{lm:crossingNEW} all use perturbations from some given policy which increase the expected value for later types while keeping earlier disclosures and experimentation choices unchanged. 
Therefore, there exists a policy satisfying the conditions in \Cref{prop:main2NEW}, which results in a Pareto improvement over transparency.
We show that the policy in \Cref{prop:main1NEW} is such a policy.

The proof of \Cref{prop:main1NEW} argues that for any policy that satisfies the necessary conditions in \Cref{prop:main2NEW}, there exists a policy that takes the simple form of \Cref{prop:main1NEW} that is a (weak) Pareto improvement over the original policy.
This implies that there exists a policy of form \Cref{prop:main1NEW} that is a Pareto improvement over transparency, too. To complete the proof, we must show that the specific policy where $\overline{t}$ is chosen to maximize social welfare is a Pareto improvement over transparency.

Suppose not, so that there exists $\hat{t}$ such that all types are weakly better off under the policy in \Cref{prop:main1NEW} with $\overline{t}=\hat{t}$ than under transparency, but that for some $t^* \ne \hat t$, some type is strictly better off under transparency than under the policy in \Cref{prop:main1NEW} with $\overline{t}=t^*$. 
We must have $t^*=T_{\overline{n}}$ in the solution of \eqref{Aeq-lemma5}. The types who invest under both $\overline{t}=\hat{t}$ and $\overline{t}=t^*$ get identical payoffs under both schemes. Hence, we focus on types that do not experiment under one scheme, which implies that they do not experiment under transparency either because any optimal $\overline{t}$ is such that $v(x(q_{\overline{t}}))<0$. 
We can bound the expected transparency payoffs of non-experimenters by using a scheme that discloses all generated good news after the last experimenting agent $\overline{q}^\text{TP}: \min\{\theta \in [0,F_n]\colon v(x(\theta))=0\}$ has experimented:\footnote{This is an improvement over transparency by the same argument as used in the proof of \Cref{prop:main2NEW}: more patient types are better off when all good news is concentrated to a one-time disclosure that leaves the marginal experimenter indifferent. In this case, good news disclosure at the end does not cause any additional delay because the no-disclosure investment payoff is not strictly below 0 after the disclosure.}
$
    e^{-r_i\t^\text{Bad}(\overline{q}^\text{TP})}G(\overline{q}^\text{TP})x_0v_G,
$
where $\t^\text{Bad}$ is the experimentation profile under transparent bad news and censored good news. We can solve for $G(\overline{q}^\text{TP})$ from $v(x(\overline{q}^\text{TP}))=0$: $
    G(\overline{q}^\text{TP})=\frac{x_0 v_G+(1-x_0)B(\overline{q}^\text{TP})v_B}{x_0 v_G}.
$
Now, $
    e^{-r_i\t^\text{Bad}(\overline{q}^\text{TP})}G(\overline{q}^\text{TP})x_0v_G,
$ becomes
\begin{align}\label{Aeq-pareto2}
    e^{-r_i\t^\text{Bad}(\overline{q}^\text{TP})}(x_0v_G+(1-x_0)B(\overline{q}^\text{TP})v_B),
\end{align}
which must be weakly smaller than what type $i$ gets under the optimal policy because a feasible deviation of investing at $\t^\text{Bad}(\overline{q}^\text{TP})$ absent bad news yields exactly \eqref{Aeq-pareto2}.
\qed 



\newpage

\linespread{1}\selectfont
\bibliography{references.bib}

 
\newpage

\linespread{1.5}\selectfont

\setcounter{page}{1}

\section{Supplemental Appendix} \label{sec:online_appendix}

In this Online appendix we show the equivalence, from the perspective of the agents, of the representations of disclosure policies in terms of families of cdfs $D^B$ and $D^G$ and the processes $H^B_t$ and $H^G_t$ as introduced in the Model Section \ref{sec:model}.

We also provide characterization for the transparency benchmark for mixed news and an illustrative example with homogeneous agents. 

\subsection{Equivalent representation of disclosure policies}\label{sec:Z-processes}

Fix any non-decreasing experimentation process $q_t$. We will denote by $\tilde D^{ \w}$, the conditional probability measure and by $D^\w$ its corresponding cdf. 
Formally, the sender commits to two (regular) conditional probabilities over disclosure delays,  $\tilde D^{ \w}:  \mathcal B(\mathbb{R}_+)\times \mathbb{R}_+  \rightarrow [0,1]$, such that for all $ A \in \mathcal B(\mathbb{R}_+)$, the function  $\tilde D^{ \w}(A \lvert \cdot )$ is Borel measurable; and  for any generation time $s \in \mathbb{R}_+$, the conditional distribution $\tilde D^{ \w}(\cdot  \lvert s)$  is a probability measure over disclosure delays  for each $\w$ in \{G,B\}.

We construct for any $\w$-evidence disclosure policy specified above, given by a family of conditional cdfs $D^\w(\cdot \lvert s)_{s\ge 0}$,  the corresponding process $\Paren{H_t^\w}_{t\ge 0}$.
Consider $\w=B$. The probability  that $B$-evidence is disclosed by (before or at) time $t$ is
 \begin{align*}
     \int_{0}^{t} D^B(t -s \lvert s) \de B(q_{s}) , 
\end{align*}
where $B(q_{s}) $ is the probability that $B$-evidence has been generated by time $s$, and, for each $s\le t$, $D^B(t -s \lvert s)$ is the probability that the piece of $
B$-evidence generated at time $s$ is disclosed no later than at time $t$.
The corresponding probability of disclosure \textit{before} any time $t$ is given by the following identity, involving the left-limit of the above probability 
\begin{align*}  H^B_t = \lim_{\e \downarrow0} \  \int_{0}^{t-\e} D^B(t-\e -s \lvert s) \de B(q_{s}) ,
\end{align*}
The same construction applies for $H^G$, replacing cdfs $D^B$ with $D^G$ and generation probabilities $B$ with $G$. 

Since cdfs $D^\w(t-s \lvert s)$ are non-decreasing in $t$ for all $s$,  the term on the right-hand side is non-decreasing in $t$. Further,  if $D^\w(t-s\lvert s) =0$ for all $s <t$, then $H^\w_t = 0$; and if $D^\w(t-s\lvert s) =1$ for all $s < t$, then if $\w=B\colon$ $ H^B_t = B(q_{t-})$, and if $\w=G\colon$ $H^G_t = G( q_{t-})$.

Conversely, for any pair of generation and disclosure processes, we can find a family of conditional distributions satisfying the relationship above. This follows from Theorem 1 (part (i) $\Longrightarrow$ (v)) in \cite{kamae1977stochastic}.
For concreteness, we provide a characterization of $D^B$ that yields $H^B$ for a given path $q$. Define $\hat{H}^B_t:=\lim_{\e \downarrow 0}H_{t+\epsilon}^B$ and let $\hat{s}(q):=\min\{t:\hat{H}_t^B\geq B(q)\}$. Now for all continuity points of $q_s$, i.e. $q_s=q_{s-}$, we set  
        \begin{align*}
            &D^B(\D|s)=\begin{cases} &0 \quad \text{ if } \D < \hat{s}(q_s)-s,
            \\ &1 \quad \text{ if } \D \geq \hat{s}(q_s)-s
            \end{cases}
        \end{align*}
    For $q_s>q_{s-}$, we set
        \begin{align*}
            &D^B(\D|s)=\min\{\max\{\frac{\hat{H}^B_{s+\D}-B(q_{s-})}{B(q_{s})-B(q_{s-})}, 0\}, 1\}
        \end{align*}
We also set $D^B(\infty, s) = 1$ for the case that $H^B_\infty < B(g_s)$. 

This characterization discloses data in the order in which they were generated.  Now, it is easy to verify that $ \int_{0}^t D_B(t -s\lvert s)   \de B(q_s)=\hat{H}_t^B$ and, hence $\lim_{\e \downarrow 0}\int_{0}^{t-\e} D_B(t-\e -s\lvert s)   \de B(q_s)={H}_t^B$.

Finally, notice that as a monotone function,  $D^B(\cdot, s)$ has only countably many discontinuity points, and hence the family $D^B$ is Borel measurable.

\subsection{Transparent benchmark}\label{Asec-TP}

Consider transparent disclosure, that is $(H^B_t,H^G_t)= (B(q_{t-}),G(q_{t-}))$   for all $t$. 
Let $q_t^\text{TP}$ denote the mass of agents who have experimented by time $t$. 

We consider first the case with joint or pure bad news with $B$ everywhere strictly increasing and then the case of pure good news to prove \Cref{lemma-GN-BM}. 
$B$ strictly increasing implies that $q_t^\text{TP}$ is continuous because otherwise there would be a strict improvement to invest after the atom of other agents has invested rather than at the same time with them. 
Recall from the discussion of the agent problem \eqref{eq:agent_prob} that agents invest in increasing order of $\th$. Thus, we can find times  $\{T _{i}\}_{i=0}^{n}$ such that all type $i$ agents invest in the interval $[T_{i-1}, T_{i})$ absent news.

 Suppose no news has arrived by some time $t\in [T_{i-1}, T _{i})$. 
In equilibrium, type $r_i$ has to be indifferent between stopping immediately at $t$  or stopping at time $t+\de t$ if no bad news has arrived during $[t,t+\de t)$.\footnote{There cannot be atoms (no strict preference to experiment) or time intervals where no one experiments (the marginal agent cannot strictly prefer waiting). If there were, then investing before the break would be strictly better than investing at the end of the break.} 
Ignoring terms of order higher than $\de t$, the expected time-$t$ payoff from the waiting an instant $\de t$ and experimenting at $t+\de t$ if no bad news arrived is
\begin{align}\label{Aeq-tp_wait}
x_t \frac{G(q^\text{TP}_{t+\de t})-G(q^\text{TP}_t)}{1-G(q_t^\text{TP})}  v_G +     e^{-r_i \de t} \left(x_t \frac{1-G(q^\text{TP}_{t+\de t})}{1-G(q^\text{TP}_t)} v_G+(1-x_t)\frac{1-B(q^\text{TP}_{t+\de t})}{1-B(q^\text{TP}_t)}v_B \right),
\end{align}
where we used  $x_{t+\de t} = x_t \frac{1-G(q^\text{TP}_{t+\de t})}{1-G(q^\text{TP}_t)} /(x_t \frac{1-G(q^\text{TP}_{t+\de t})}{1-G(q^\text{TP}_t)} +(1-x_t)\frac{1-B(q^\text{TP}_{t+\de t})}{1-B(q^\text{TP}_t)})$. The event when bad news arrives during $\de t$ does not show up in \eqref{Aeq-tp_wait} as it yields zero payoffs.
The payoff from waiting must equal $v(x_t)$, the expected payoff from investing at $t$ immediately. Forming the Taylor approximation and taking the limit as $\de t\to0$ from that indifference condition yields 
\begin{align}\label{eq-BM_q} 
  r(q^\text{TP}_t) v(x_t)=  (1-x_t) \frac{b (q^\text{TP}_t)\dot q^\text{TP}_t}{1-B(q^\text{TP}_t)}  (-v_B)  . 
\end{align} 
This is a first-order ordinary differential equation (ODE) where $x_t$ is itself a  function of $q^\text{TP}_t$. It does not have  a closed form solution for general $B$ and $G$.

Experimentation ends when $q^\text{TP}_t = F_n$ or at the first $q$ such that $x(q)\leq x^{myop}$.
   \begin{figure}[ht]
  \centering
\begin{tabular}{p{0.45\textwidth}p{0.45\textwidth}}
\vspace{0pt}  
\begin{tikzpicture}
\begin{axis}[
                        axis lines = center,
                        scale = 0.75,
                        xtick = {0.001, .245,.45},
                        xticklabels = {{\scriptsize $0$},{ \scriptsize $T_1$},{\scriptsize $T_2$}},
                        ytick = {0.001,.5,1},
                        yticklabels = {{\scriptsize $0$},{ \scriptsize $F_1$},{\scriptsize $F_2$}},
                        xmin = 0,
                        xmax = .5,
                        ymin = 0,
                        ymax = 1.2, 
                        xlabel = {$t$},
                        ylabel = {$q^\text{TP}$
                        },
                        x label style={at={(current axis.right of origin)},anchor=west},    
                        y label style={at={(current axis.above origin)},anchor=south},    
                    ]
                    
                 \addplot [dashed, gray, domain=0:.45, samples=2]{1} ;   
                    \addplot [dashed, gray, domain=0:.245, samples=2]{.5} ;

                 \draw[->, orange, >=stealth, dotted, thick] (axis cs:0.15,0.215) to [bend left] (axis cs:0.148,.985);
                 \draw[->, orange, >=stealth, dotted, thick] (axis cs:0.225,0.42) to [bend left] (axis cs:0.225,.985);
                 \draw[->, orange, >=stealth, dotted, thick] (axis cs:0.3,0.591) to [bend left] (axis cs:0.3,.985);
                 
                 
                 \draw[orange, dashed, thick] (axis cs:0.15,1) -- (axis cs:0.5,1);

                 \draw[-, dashed, thick, blue] (axis cs:0.15,0.215) to  (axis cs:0.495,0.215);

                                                            \draw[-, dashed, thick, blue] (axis cs:0.225,0.42) to  (axis cs:0.49,0.42);
                              
                                \draw[-, dashed, thick, blue] (axis cs:0.3,0.591) to  (axis cs:0.495,0.591);

               \addplot [very  thick,  domain=0:.245, samples=50]{1/6*(x*3-3*ln(2*exp(3*x/3)-exp(3*x))}
                ;
             
                    \addplot [ very thick,  domain=.245:.445,samples=50]{ 1/6*(.735+x-0.245-3*ln(2*exp((.735+x-0.245)/3)-exp(.735+x-0.245)))}
                        ;

                      \addplot [very thick, domain=.445:.5,samples=2]{1};

                    \end{axis}
		\end{tikzpicture}   & \vspace{0pt} %
 \begin{tikzpicture}
	 \begin{axis}[
                        axis lines = center,
                        scale = 0.75,
                        xtick = {0.001},
                        xticklabels = {{\scriptsize $0$}},
                        ytick = {0.001,.4,.5,1},
                        yticklabels = {{\scriptsize $0$},{\scriptsize $q^\text{TP}_\infty$},{ \scriptsize $F_1$},{\scriptsize $F_2$}},
                        xmin = 0,
                        xmax = .5,
                        ymin = 0,
                        ymax = 1.2, 
                        xlabel = {$t$},
                        ylabel = {$q^\text{TP}$
                   },
                        x label style={at={(current axis.right of origin)},anchor=west},    
                        y label style={at={(current axis.above origin)},anchor=south},    
                    ]
                    
                \addplot [dashed, gray, domain=0:.5, samples=2]{1} ;   
                    \addplot [dashed, gray, domain=0:.5, samples=2]{.5} ;   
                       \addplot [dashed, gray, domain=0:.5, samples=2]{.4} ;   

   \draw[->, orange, >=stealth, dotted, thick] (axis cs:0.07,0.15) to [bend left] (axis cs:0.07,.985);
    \draw[->, orange, >=stealth, dotted, thick] (axis cs:0.17,0.26) to [bend left] (axis cs:0.168,.985);
    \draw[->, orange, >=stealth, dotted, thick] (axis cs:0.27,.315) to [bend left] (axis cs:0.27,.985);
    \draw[orange, dashed, thick] (axis cs:0.05,1) -- (axis cs:0.5,1);
            
                \draw[-, dashed, thick, blue] (axis cs:0.07,0.15) to  (axis cs:0.495,0.15);
                \draw[-, dashed, thick, blue] (axis cs:0.17,0.26) to  (axis cs:0.495,0.26);
               \draw[-, dashed, thick, blue] (axis cs:0.27,0.313) to  (axis cs:0.49,0.313);

      \addplot [ very thick, domain=0:.5, samples=50]{10*x/(20*x+3)}
                ;

              \end{axis}
		\end{tikzpicture} 
\end{tabular}  
\caption{\linespread{1.0}\selectfont \small Investment process under transparency and exponential learning with $\l_G<\l_B$ (left) and with $\l_G>\l_B$ (right). The solid black curve depicts the mass of agents who have experimented. At all times, positive or negative evidence is generated and disclosed at a positive rate. If good evidence arrives, all agents invest immediately (orange arrows), if bad evidence arrives, no additional agents invest (blue lines). }
\label{fig:Transparency}
\end{figure}

Figure \ref{fig:Transparency} depicts the experimentation dynamics under transparent disclosure when $n=2$ and exponential learning, with constant hazard rates $\lambda_B$ and $\lambda_G$. In the left plot, no news is good news: $\lambda_B>\lambda_G$. Between times $0$ and $T_1$, the impatient types $r_1$ experiment. 
At time $T_1$, all agents with type $r_1$ have invested and the more patient agents with type $r_2$ start investing. Due to the lower cost of waiting ($r_2<r_1$), the investment rate of the patient agents is initially lower and increases as the belief increases further. 
In the right plot, no news is bad news, so $\dot q^\text{TP}$ is decreasing. In this example, $F_1$ is so large that $v(x(F_1))< 0$. 
 Then, $\dot q^\text{TP}$ approaches 0 before all $r_1$-agents have experimented as the left-hand side of \eqref{eq-BM_q} approaches 0 as $x_t$ approaches $x^\text{myop}$. 
Hence, $T_1 = \infty$ and the final amount of experimentation approaches $q_{\infty}^{\text{TP}}<F_1$ that solves $v(x( q^\text{TP}_{\infty}))= 0$. 
This amount is inefficiently low. Agents disregard the social benefit from evidence generation and thus would never experiment when the expected return is negative.

Second, consider the pure good news case, i.e., suppose  that $B(q)$ is constantly 0. Then, an atom of size  $\hat q$ experiments immediately at $t=0$ since $v(x_0)>0$; experimenting gives a strictly larger payoff than waiting when there is no bad news generation: \eqref{Aeq-tp_wait} is strictly smaller than $v(x_0)$.

\subsection{Illustrative example with homogeneous agents}\label{sec-example}

 Here we illustrate that the main findings of the main paper do not hinge on discounting heterogeneity. To illustrate if and how the designer can speed up information generation by delaying disclosure, consider a simple example. Assume here that $F_n =1$ and let $n=1$ so that all agents share the same discount rate $r_1 = r$. Let good news and bad news be equally likely, $B(q) =G(q) = 1- e^{- \l q}$ for $\l >0$. In this case, under transparency, the belief is constant and the adoption process is linear: $q^\text{TP}_t=\frac{r}{\lambda}\frac{x_0v_G+(1-x_0)v_B}{(1-x_0)(-v_B)}t$.

 For illustration, consider a designer whose objective is to minimize the time at which all potential evidence has been revealed, i.e. find the smallest $T$ such that $(H^B_T,H^G_T)=(B(1),G(1))$. 
The example introduces some of our results and the way we approach the designer's problem in a simplified environment.

In this example with identical agents, a disclosure process $(H^B,H^G)$ is feasible if and only if agents are willing to invest at time 0 (as well as $H^\w_t \le B(1)=G(1)$ for all $t$):
\begin{align}\label{eq-E_IC} 
    x_0 v_G +(1-x_0) v_B \geq& \int_0^{t} e^{-r s} x_0 v_G \de H^G_s \nonumber
   \\&+ e^{-rt}
    \Paren{x_0 v_G (1-H^G_t) + (1-x_0)v_B (1-H^B_t)}, \quad \text{ for all } t\ge 0.
\end{align}
 In particular, it is necessary for information generation that agents are willing to invest at time $0$ rather than waiting until time $T$:
\begin{dense}
\begin{align}\label{eq-E_IC3}
   & x_0 v_G +(1-x_0) v_B \geq \int_0^T e^{-r s} x_0 v_G \de H^G_s  + e^{-rT}(x_0 (1-G(1)) v_G  + (1-x_0)(1-B(1))v_B ).
\end{align}
\end{dense}
Since $H^G$ must satisfy $\int_0^T e^{-r s} \de H^G_s \ge e^{-r T} H^G_T$ and $H^G_T=G(1)$ by definition, the inequality above gives a lower bound $T^*$ for the disclosure time $T$:
\begin{align}\label{eq-E_IC2}
   & e^{-rT}\leq \frac{x_0 v_G +(1-x_0) v_B}{x_0 v_G +(1-B(1))(1-x_0) v_B} =:e^{-rT^*}.
\end{align}

Consider a disclosure policy where the designer commits to hide all news until time $T^*$: ${H}_t=(0,0)$ for all $t<T^*$ and ${H}_{T^*}=(B(1),G(1))$. Then, the constraint \eqref{eq-E_IC} is trivially satisfied for all $t< T^*$ and holds as an equality for $t=T^*$. Hence, the designer can implement the lower bound $T^*$ by postponing all disclosure.

Is fully postponing all disclosure necessary for implementation? Consider good news first. 
 Observe that $\int_0^{T^*} e^{-r s} \de H^G_s >  e^{-r T^*} H^G_{T^*}$ if $H_t^G>0$ for any $t<T^*$. This then implies that the right-hand side of the necessary condition \eqref{eq-E_IC3} is strictly above $e^{-rT^*}( x_0 v_G +(1-B(1))(1-x_0) v_B)$, making the disclosure process infeasible, independent of the timing of bad news. Therefore, we conclude:
\begin{observation}\label{claim-GN}
  Any disclosure policy that implements $H_{T^*}=(B(1),G(1))$ never discloses good evidence strictly before $T^*$: \ ${H}_t^G=0$ for all $t<T^*$.
\end{observation}

Next, consider bad news. 
Fixing $H^B_T=B(1)$, the exact disclosure time of negative evidence before $T$ does not affect the necessary condition \eqref{eq-E_IC3}. 
However, if the disclosure is too fast, the constraint \eqref{eq-E_IC} may be violated for earlier times $t<T$.
It is helpful to consider what happens if the designer discloses all bad news immediately: we say that an information policy has \textit{transparent breakdowns} if   $H_t^B=B(q_{t-})$ for all $t$. However, the evolution of $q$ depends on the good news policy and may hence differ from the transparent benchmark $q^\text{TP}_t$. 

Because we are interested in implementing full revelation at $T^*$ and since we know by Observation \ref{claim-GN} that it requires hiding good news until $T^*$,  consider transparent breakdowns under the assumption that good evidence is disclosed only at time $T^*$. 
The equilibrium investment process $q$ evolves such that all agents remain indifferent between experimenting and waiting. The indifference condition takes the same form as \eqref{eq-BM_q} under the fully transparent benchmark.\footnote{With homogeneous agents ($r$ instead of $r_i$) and with $\lambda_B=\lambda$.} 
In contrast to \eqref{eq-BM_q}, however, the belief is now $x_t=x_0/(x_0+(1-x_0)e^{-\lambda q_t})$ because good evidence is not disclosed. 
Because $x_t>x_0$ for all $t\in(0,T^*)$, the investment $q$ evolves faster than under full transparency: hiding good news makes agents willing to experiment earlier.\footnote{To see this note that the left-hand side of indifference condition \eqref{eq-BM_q} is decreasing in $x_t$ and the right-hand side is increasing. When the agent believes that the bad state is less likely he expects less negative evidence, decreasing the benefit of waiting; and his expected investment return is higher, increasing the cost of waiting.} 
The investment amount $q_t$ that solves the indifference condition with censored good news reaches 1 exactly at $t=T^*$. This confirms that transparent breakdowns implement full revelation at $T^*$.
     Because agents are indifferent at all $t\leq T^*$, the policy with transparent breakdowns also maximizes information revelation prior to time $T^*$, conditional on full revelation by time $T^*$:
\begin{observation} \label{Obs:BN}
      Among all disclosure policies that implement $H_{T^*}=(B(1),G(1))$, transparent breakdowns, i.e. $ H^B_t = B(q_{t-})$ for all $t$,  maximizes ${H}_t^B$ at all times $t$.
\end{observation}

We find that the designer who seeks to maximize information revelation hides good evidence but reveals bad evidence. These observations are shared with the main model where the designer seeks to maximize the expected welfare of the agents under discounting heterogeneity. With homogeneous agents, the disclosure policy has no effect on welfare. Sharing the same discount rate, every agent gets the same expected payoff as the first investor. 
Observation \ref{Obs:BN} shows that censoring earlier bad evidence does not allow the designer to disclose more cumulative bad evidence at later times.\footnote{Note that this \textit{does} work with good evidence. By Observation \ref{claim-GN}, the maximal feasible value of $H^G_{T^*}$ is strictly increased by delaying good news prior to $T^*$. 
}

\end{document}